\documentclass[12pt]{article}

\usepackage{amsmath,amssymb,amsfonts,epsfig,cite,setspace,bigstrut,framed}
\usepackage[all]{xy}
\usepackage{color}
\usepackage{pifont}

\makeatletter \@addtoreset{equation}{section} \makeatother
\renewcommand{\theequation}{\thesection.\arabic{equation}}

\begin{document}

\vskip 0.25in

\newcommand{\todo}[1]{{\bf\color{blue} !! #1 !!}\marginpar{\color{blue}$\Longleftarrow$}}
\newcommand{\nn}{\nonumber}
\newcommand{\comment}[1]{}
\newcommand\T{\rule{0pt}{2.6ex}}
\newcommand\B{\rule[-1.2ex]{0pt}{0pt}}

\newcommand{\cM}{{\cal M}}
\newcommand{\cW}{{\cal W}}
\newcommand{\cN}{{\cal N}}
\newcommand{\cH}{{\cal H}}
\newcommand{\cK}{{\cal K}}
\newcommand{\cT}{{\cal T}}
\newcommand{\cZ}{{\cal Z}}
\newcommand{\cO}{{\cal O}}
\newcommand{\cQ}{{\cal Q}}
\newcommand{\cB}{{\cal B}}
\newcommand{\cC}{{\cal C}}
\newcommand{\cD}{{\cal D}}
\newcommand{\cE}{{\cal E}}
\newcommand{\cF}{{\cal F}}
\newcommand{\cX}{{\cal X}}
\newcommand{\IA}{\mathbb{A}}
\newcommand{\IP}{\mathbb{P}}
\newcommand{\IQ}{\mathbb{Q}}
\newcommand{\IH}{\mathbb{H}}
\newcommand{\IR}{\mathbb{R}}
\newcommand{\IC}{\mathbb{C}}
\newcommand{\IF}{\mathbb{F}}
\newcommand{\IV}{\mathbb{V}}
\newcommand{\II}{\mathbb{I}}
\newcommand{\IZ}{\mathbb{Z}}
\newcommand{\re}{{\rm Re}}
\newcommand{\im}{{\rm Im}}
\newcommand{\tr}{\mathop{\rm Tr}}
\newcommand{\ch}{{\rm ch}}
\newcommand{\rk}{{\rm rk}}
\newcommand{\ext}{{\rm Ext}}
\newcommand{\bi}{\begin{itemize}}
\newcommand{\ei}{\end{itemize}}
\newcommand{\beq}{\begin{equation}}
\newcommand{\eeq}{\end{equation}}
\newcommand{\bea}{\begin{eqnarray}}
\newcommand{\eea}{\end{eqnarray}}
\newcommand{\ba}{\begin{array}}
\newcommand{\ea}{\end{array}}

\newcommand{\CA}{\mathbb A}
\newcommand{\CP}{\mathbb P}
\newcommand{\tmat}[1]{{\tiny \left(\begin{matrix} #1 \end{matrix}\right)}}
\newcommand{\mat}[1]{\left(\begin{matrix} #1 \end{matrix}\right)}
\newcommand{\diff}[2]{\frac{\partial #1}{\partial #2}}
\newcommand{\gen}[1]{\langle #1 \rangle}

\newtheorem{theorem}{\bf THEOREM}
\newtheorem{proposition}{\bf PROPOSITION}
\newtheorem{observation}{\bf OBSERVATION}

\def\theequation{\thesection.\arabic{equation}}
\newcommand{\setall}{
	\setcounter{equation}{0}
}
\renewcommand{\thefootnote}{\fnsymbol{footnote}}

\begin{titlepage}
\vfill
\begin{flushright}
{\tt\normalsize KIAS-P15014}\\

\end{flushright}
\vfill

\begin{center}
{\Large\bf Mutation, Witten Index, and Quiver Invariant}

\vskip 1.5cm

Heeyeon Kim\footnote{\tt heeyeon.kim@perimeterinstitute.ca},
Seung-Joo Lee\footnote{\tt seungsm@vt.edu}, and
Piljin Yi\footnote{\tt piljin@kias.re.kr}
\vskip 5mm
{\it $^*$Perimeter Institute for Theoretical Physics, \\31 Caroline Street North, Waterloo, N2L 2Y5, Ontario, Canada}
\vskip 3mm
{\it $^\dagger$Department of Physics, Robeson Hall, Virginia Tech, \\
Blacksburg, VA 24061, U.S.A.
}
\vskip 3mm
{\it $^\ddagger$School of Physics, Korea Institute
for Advanced Study, Seoul 130-722, Korea}

\end{center}
\vfill

\begin{abstract}
We explore Seiberg-like dualities, or mutations, for ${\cal N}=4$ quiver
quantum mechanics in the context of wall-crossing. In contrast to higher
dimensions, the 1d Seiberg-duality must be performed with much care.
With fixed Fayet-Iliopoulos constants, at most two nodes can be mutated,
one left and the other right, mapping a chamber of a quiver into a
chamber of a mutated quiver. We delineate this complex pattern for triangle
quivers and show how the Witten indices are preserved under such finely
chosen mutations. On the other hand, the quiver invariants, or
wall-crossing-safe part of supersymmetric spectra, mutate
more straightforwardly, whereby a quiver is mapped to a quiver.
The mutation rule that preserves the quiver invariant is different
from the usual one, however, which we explore and confirm numerically.

\end{abstract}

\vfill
\end{titlepage}

\tableofcontents
\renewcommand{\thefootnote}{\#\arabic{footnote}}
\setcounter{footnote}{0}
\vskip 2cm

\section{Introduction}

Quiver quantum mechanics, and more generally gauged linear sigma models
quantum mechanics (1d GLSM)
with four or less supersymmetries, exhibit wall-crossing behavior
where the Witten index jumps discontinuously under continuous
deformation of Fayet-Iliopoulos constants $\zeta$. With ${\cal N}=4$
supersymmetry, this is directly connected to the wall-crossing
of Seiberg-Witten theories \cite{Seiberg:1994rs,Seiberg:1994aj,Ferrari:1996sv}
via Calabi-Yau compactification of type II string theories.
Such
discontinuities have been studied in many different approaches in the past.
The fundamental mechanism of BPS state disappearance in 4d was
understood fairly early via multi-center nature of generic BPS states
\cite{Lee:1998nv} which was followed by explicit state
counting and wall-crossing of multi-center BPS bound states
in the weakly coupled regime of rank two or higher gauge theories
\cite{Bak:1999da,Gauntlett:1999vc, Stern:2000ie} and then later
in the supergravity or Calabi-Yau setting \cite{Denef:2000nb,Denef,Denef:2007vg}.

These early works inspired two different approaches to the general
wall-crossing problems. One resorted to more mathematical
reformulation, culminating in the Kontsevich-Soibelman wall-crossing
formulae \cite{KS,GMN1}. This is suitable for noncompact
Calabi-Yau examples, e.g., Seiberg-Witten theories, and was later further
clarified and expanded via compactified (2,0) theory \cite{GMN2,Gaiotto:2010be}.
The other, more faithful to the physical picture of multi-center
BPS states, was developed by and large parallel to the former.
The latter resulted in a very comprehensive and universal index formulae
\cite{deBoer:2008zn,Manschot:2010qz,Manschot:2011xc,Kim:2011sc},
from which wall-crossing formulae followed. The latter, when
restricted appropriately to fit the smaller scope of the former,
has been shown to be solutions to the Kontsevich-Soibelman
wall-crossing algebra \cite{Sen:2011aa}.

Despite such a long history and several breakthroughs, there are
some important  questions remaining. For example, while we have several different
wall-crossing formulae and index formulae, actual evaluation of them
in examples beyond 4d rank one theories are hardly understood at a
systematic level. Also, beyond such more technical issues, there is
also a conceptual mystery surrounding part of supersymmetric spectra
that remain robust across walls of marginal stability. In the Calabi-Yau
setting, this question appears to be essential to complete classification
and counting of supersymmetric cycles of {\it compact} Calabi-Yau three-folds,
and in particular to microstate counting of 4d ${\cal N}=2$ BPS black holes.

To explain these wall-crossing-safe states, let us come back
to 1d quiver quantum mechanics,
which are low energy dynamics of D3-branes wrapped on special Lagrange
cycles of Calabi-Yau three-fold \cite{Kachru,Denef}. Such quiver quantum
mechanics has resurfaced more directly from Seiberg-Witten theories
as well; One recent is via low energy dynamics of BPS solitons in
strongly coupled  regimes \cite{Lee:2011ph,Kim:2011sc}, while another
is from realization of Seiberg-Witten theory as (2,0) theories
compactified on Riemannian surfaces with punctures \cite{GMN2,Alim:2011ae,Alim:2011kw}.
Wall-crossing of 4d BPS states then translates to appearance and
disappearance of supersymmetric vacua of such 1d quiver theories.

For a simpler class of quivers like two-node Kronecker quivers,
all supersymmetric vacua disappear simultaneously across a single
wall at $\zeta=0$.
Whenever a quiver comes with a superpotential, however, a subtlety arises.
Spectrum is split into part that disappear at such walls  and part
that remain robust everywhere in FI constant space
\cite{Denef:2007vg,Bena:2012hf,Lee:2012sc,Lee:2012naa}.
As we will review in a later section, the latter states are
all angular momentum, or $SU(2)_R$ singlets \cite{Lee:2012sc,Lee:2012naa},
and serve as building blocks in the multi-center picture
or the Coulombic picture, of the wall-crossing \cite{Manschot:2013sya,Manschot:2014fua}.
The latter class of states should exist even when all FI constants
are set to zero, and have been dubbed the {\it quiver invariant} \cite{Lee:2012sc,Lee:2012naa},
for an obvious reason; these states, or part of Witten index that
captures them, are invariant properties of the quiver itself
rather than those of individual chambers with distinct Witten indices.

Existence of such wall-crossing-safe states implies that
the wall-crossing formulae are nowhere
enough for counting ground states of quiver quantum
mechanics, or equivalently counting BPS spectra in
four dimensions. For the Kontsevich-Soibelman wall-crossing
formulae, in fact, the quiver invariants should be regarded
as input data rather than solution to their algebraic constraint.
In 4d context, the quiver  invariant seems to count degeneracy
of single-center ${\cal N}=2$ BPS black holes \cite{Lee:2012naa},
which also tells us that for black hole microstate counting,
it is not the wall-crossing pattern that matters but rather
one must compute Witten indices and the quiver invariants
more directly. Beyond simple cases like $SU(2)$ Seiberg-Witten,
therefore, the need for direct counting of Witten indices
is all the more pressing.

Equivariant Witten index counting for general ${\cal N}\ge 2$
gauged quantum mechanics has been established in a recent
work \cite{HKY}, where the wall-crossing in $\zeta$ space
is also captured and accounted for correctly.
Although actual evaluation, some $r$-dimensional contour
integrals, is riddled with subtleties and also mired by heavy
computational cost as rank $r$ grows, this result, to which we will refer
as HKY, represents the most comprehensive approach to counting
supersymmetric ground states of 1d GLSM. It represents the first
systematic and comprehensive counting method;
although there had been systematic approaches such as that of
Reineke \cite{Reineke} or those based on Coulombic approximation
\cite{Manschot:2010qz,Kim:2011sc}, these are effectively restricted
to quivers without superpotentials and  in particular
cannot count black hole microstates by themselves.
This new approach supersedes existing geometrical methods
such as the Abelianization scheme reviewed in Ref.~\cite{Lee:2013yka},
and has been used for various nontrivial examples. For low-rank or
Abelian examples, the prescription  is very effective and Witten indices
have been computed for many ${\cal N}=2,4$ GLSM's.

Computation of high rank non-Abelian GLSM's, other than some of
very simple quivers, remains technically challenging,
however.\footnote{For the simple Kronecker quivers,
the large-rank scaling  behavior has been also obtained with help
of this approach \cite{Cordova:2015qka,Kim:2015oxa,Cordova:2015zra},
although the scaling behavior found here is not related to that
of ${\cal N}=2$ BPS black holes but rather intrinsic to rank 2
or higher field theory BPS states.} On the other hand, a very
suggestive scaling behavior with growing intersection numbers
has been seen in the wall-crossing-safe part of spectrum
\cite{Denef:2007vg,Bena:2012hf,Lee:2011ph}. This could be related to
Witten indices of high rank quivers in two possible ways. One is via the
MPS expansion which expresses the index of a high rank quiver
via a partition sum of the rank vector where high rank often
translates to the high intersection numbers in the computational
middle steps. Another possibility is
the so-called mutation map, which can preserve Witten index
under favorable circumstances while mixing up rank vectors and
intersection numbers.

The mutation, which is a form of Seiberg-duality for the
quiver quantum mechanics, has been very successfully used
for obtaining BPS spectra of rank-one Seiberg-Witten
theories by  Alim et.al. \cite{Alim:2011kw} who
argued how two different-looking quivers, with very
different ranks, can possess chambers of the same Witten indices
and explained how two such can be viewed as a mere change
of basis. The basis element in question can be either
a specific set of simple dyons for Seiberg-Witten theory, or
a set of special Lagrangian submanifolds for Calabi-Yau three-fold.
Thus, one immediate problem is to verify the proposed mutation
invariance against explicit Witten index counting.
Because the mutation always acts on a single node at a time
and transforms the adjacent nodes by the connecting arrows, the simplest
prototype where all the subtleties of mutation can be seen
is the cyclic triangle quivers where each node
is connected to a pair of nodes each with
ingoing and outgoing arrows. One main objective of this note
is to study this class of quivers in detail and demonstrate
how mutation map manifests in HKY's Witten index counting.

This mutation map is, however, rather specific in that
it requires certain inequalities among FI constants, $\zeta$.
Because of this,  the map cannot map all physical chambers
of a quiver to those of one mutated quiver.
Chamber by chamber, allowed mutations are generically all
different. While the mutation
can represent a powerful method for relating quivers of
different ranks and intersection  numbers, this severe
$\zeta$-dependence is subtle enough to hinder most practical
applications generally. On the other hand, such  subtleties
turn out to be absent as far as quiver invariants go.
As noted above, for general quivers that accept superpotential,
the notion of the quiver invariant has emerged as key ingredient
to understanding of the spectra
\cite{Lee:2012sc,Lee:2012naa,Manschot:2013sya,Lee:2013yka,Manschot:2014fua}.
Because the quiver invariant is a basic property of a quiver,
independent of chamber choices, we can anticipate that
the mutation rule preserving quiver invariant, if it exists
at all, should not be mired by FI constants. However, the usual
mutation rule that preserves Witten index chamber-wise is
clearly inadequate for this as one immediately sees
counterexamples where the mutated quiver and the original
quiver have two very different chamber structures.

It turns out that the relevant mutation rule for the quiver
invariant is identical to the usual mutation rule, except
that it shifts the rank of the mutating node differently
as
\begin{equation}
N_k\quad\rightarrow\quad  -N_k + {\rm min}\left(N_f^{(k)},N_a^{(k)}\right)\ ,
\end{equation}
where $N_{f}^{(k)}$ and  $N_{a}^{(k)}$ are, respectively,
the total number of chiral fields in the fundamental representations
and the total number of chiral fields in the anti-fundamental
representations, with respect to $U(N_k)$.
This action is different from the usual mutation rule, yet
preserves the quiver invariants. Because of the
chamber-independent nature of the quiver invariant,
this mutation on quiver invariant can act on any node
of the quiver, regardless of $\zeta$ values.

In section 2, we overview the quiver data and the quiver
mutations. Here we will introduce a few manifestations
of mutation maps with different action on ranks but with
a common action on adjacency matrix. One of them, to be
distinctly denoted as $\tilde\mu$, will turn out to be
the right action that preserves the quiver invariant.
Section 3 is devoted to a brief review of HKY index
formulae for gauge quantum mechanics, which is our
main tool for checking how mutation acts on Witten indices
and quiver invariants.
After a review of wall crossing, Witten index, and
quiver invariant in section 4, we move in section 5 to
ordinary mutations $\mu$ and test how they preserve Witten indices
selectively, using HKY's Witten index formulae.
Section 6 discusses mutation on quiver invariant,
given by the alternate action $\tilde\mu$, and makes
predictions for several sequences of triangle quivers,
numerical confirmations of which can be found in Appendix~\ref{appA}.

\section{Quivers and Quiver Mutations}

The quiver mutation rule takes a supersymmetric quiver theory
with four supercharges and maps it to another such theory
with different gauge group and matter content. More specifically,
a quiver theory is specified by the following set of data:

\begin{itemize}

\item
The nodes, labeled by $i$, with ranks $N_i$. Each node represents
a vector multiplet with the gauge group $U(N_i)$.

\item
The adjacency matrix, $b=\left[ b_{ij} \right]$, which counts the arrows
from node $i$ to node $j$. Positive $b_{ij}$ counts
the chirals in the bifundamental representation, $(\bar{N}_i,{N}_j)$.

\item
Fayet-Iliopoulos (FI) constant, $\zeta_i$, for each node. For this note
we take the normalization for $\zeta$'s such that FI term in the
Lagrangian is of the form
$$-\zeta_i\int dt \;{\rm tr}D_i\ ,$$
where $D_i$ is the auxiliary field in the gauge multiplet of $U(N_i)$.
\item
$R$-charge assignment $R_{ij}$ for chiral multiplets.

\end{itemize}

Recall that the quiver quantum mechanics would be a low energy
dynamics of BPS state of total charge $\Gamma=\sum_i N_i\gamma_i$,
of some 4d $\cN=2$ theories \cite{Denef}. The simplest setting where quiver
quantum mechanics emerge is type IIB theory
compactified on a Calabi-Yau three-fold. The effective theory in the remaining
four dimensions carries ${\cal N}=2$ supersymmetry, and the BPS states thereof are
realized as D3-branes wrapped on special Lagrange subcycles of the Calabi-Yau.
When the cycle is rigid, as with $S^3$, the vector multiplet on the D3-brane
reduces to quantum mechanical vector multiplet whose content is the same
as ${\cal N}=1$ vector multiplet in four dimensions.\footnote{When the special
Lagrange cycle is not rigid, there could
be further chiral multiplets, such as in the adjoint representation, although
in this note we won't consider such cases.} We denote the bosonic
part of the multiplet as
$$(A_0,x_1,x_2,x_3)\ ,$$
where the latter three transform under $SU(2)_R$ $R$-symmetry as triplet.
Generators of this $SU(2)_R $ are denoted as  $J_{1,2,3}$. In addition
there is also $U(1)_R$ symmetry which is inherited from its four-dimensional
reincarnation. We denote its half-integral generator by $I$.

When we view the quiver theory as the dynamics of D-branes wrapped
on supersymmetric cycles in a Calabi-Yau three-fold, with the charge label $\gamma$'s, the adjacency matrix, $b=\left[ b_{ij}\right]$, of the quiver counts their intersections as $b_{ij} =\langle \gamma_i,\gamma_j\rangle$,
whereby $b$ is manifestly an antisymmetric matrix. Finally, $\zeta_i$ is
related to the phase of the central charge of the cycle $\gamma_i$.

The quiver mutation maps a quiver ${\rm Q}=(N;b)_\zeta^R$ to another quiver
$\widehat {\rm Q}=(\widehat{N};\hat b)_{\hat \zeta}^{\hat R} $. Mathematical literatures usually start
with mutation rule for the matrix $b$, but for our purpose it is
more transparent to start with mutation of the underlying charges
$\gamma_i$. For each node, say, for node $k$, one can define two different mutation maps
$\mu_k^{L,R}$ which can be understood most easily via their
action on $\gamma_i$'s. For the left mutation on node $k$, we have
\begin{equation}
\mu_k^{L}(\gamma_i)= \left(\begin{array}{ccl}
-\gamma_k & \qquad & i=k \\ \\
\gamma_i+ [b_{ki}]_+ \gamma_k && \hbox{otherwise}
\end{array}\right.
\end{equation}
where $[b]_+$ is $b$ for positive $b$ and zero otherwise.
The right mutation is a mirror image of this,
\begin{equation}
\mu_k^{R}(\gamma_i)= \left(\begin{array}{ccl}
-\gamma_k & \qquad & i=k \\ \\
\gamma_i+ [b_{ik}]_+ \gamma_k && \hbox{otherwise}
\end{array}\right.
\end{equation}
Giving the mutation rule to $\gamma$'s first has the
advantage that the rule on $\zeta$ follows automatically as,
\begin{equation}
\mu_k^{L}(\zeta_i)= \left(\begin{array}{ccl}
-\zeta_k  & \qquad & i=k \\ \\
\zeta_i+ [b_{ki}]_+ \zeta_k  && \hbox{otherwise}
\end{array}\right.
\end{equation}
and
\begin{equation}
\mu_k^{R}(\zeta_i)= \left(\begin{array}{ccl}
-\zeta_k  & \qquad & i=k \\ \\
\zeta_i+ [b_{ik}]_+ \zeta_k  && \hbox{otherwise}
\end{array}\right.
\end{equation}
Both of these two mutations on $\gamma_i$'s lead to
a common rule for $b$ as
\begin{equation}
\mu_k(b_{ij})= \left(\begin{array}{ccl}
-b_{ij} & \qquad & \hbox{if } i=k \hbox{ or } j=k \\ \\
b_{ij} + {\rm sgn}(b_{ik})[b_{ik}b_{kj}]_+ && \hbox{otherwise}
\end{array}\right.
\end{equation}
where we dropped the superscript since the left and the right mutations
lead to a common rule. This common rule on the adjacency matrix is
the usual starting point for the cluster algebra. Shift of $R$-charges,
$\mu_k^{L,R}(R_{ij})$, is somewhat ambiguous, due to possible mixing with
gauge and flavor charges, some aspects of which will be discussed in
Section 5.

\begin{figure}[t]
\centering
\includegraphics[width=0.8\textwidth]{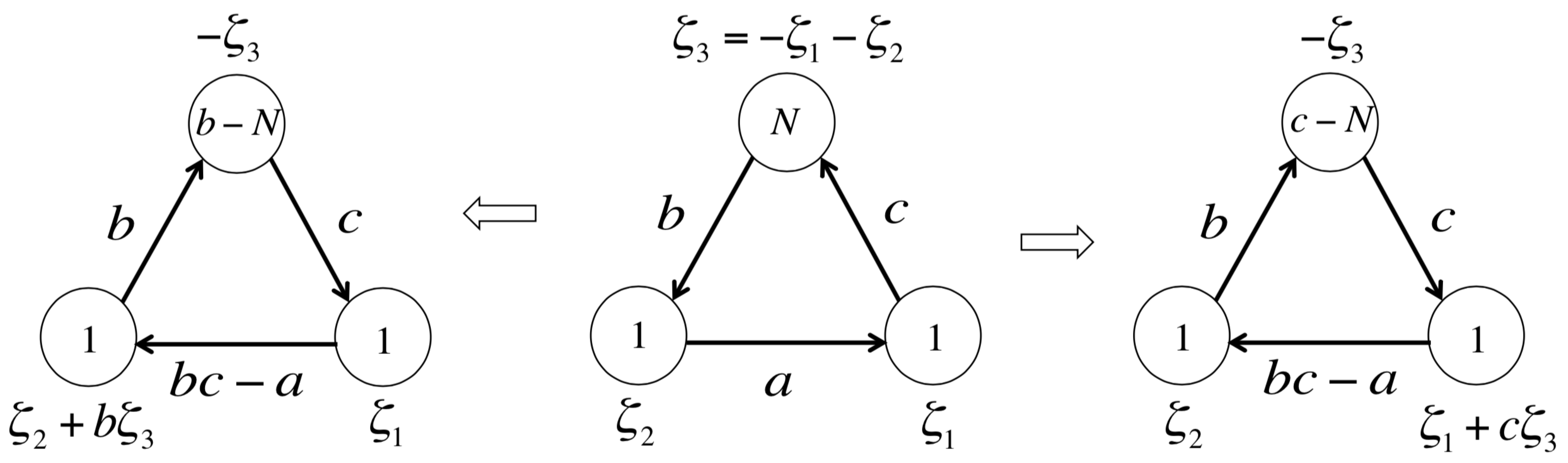}
\caption{\small   The left and the right mutations $\mu^{L,R}_3$ on node 3
for $(1,1,N)$ quivers. The integers inside circles
are ranks, while the FI constants are displayed next to them.}
\label{LR}
\end{figure}

When we try to apply the above mutation rule to quivers with loops,
it is important to restrict to the set of quivers without 1-cycles nor 2-cycles,
where the 1-cycles refer to arrows start and end at the same node,
and the 2-cycles refer to two \emph{non-canceling} arrows
with opposite direction between two nodes. Also,
the superpotential is assumed
to be generic but consistent with the gauge symmetry and $R$-symmetry.
The latter implies that $W$ is of charge 2 with respect to the
$U(1)_R$ in the convention where $R$-charges of supercharges are $\pm 1$.
One underlying assumption in the above is that we pair-annihilated
chirals of mutual charge conjugate by assigning appropriate $R$-charges
to them to allow for a bilinear term in $W$, which lifts them pairwise from the low
energy dynamics.

Finally, the mutated quiver needs the rank data $\widehat{N}_i=\mu_k(N_i)$.
One natural prescription is to keep $\Gamma\equiv \sum_i N_i\gamma_i$
invariant under the mutation, for which we have
\begin{equation}
\mu_k^{L}(N_i)= \left(\begin{array}{ccl}
-N_k + \sum_j [b_{kj}]_+ N_j & \qquad & i=k \\ \\
N_i  && \hbox{otherwise}
\end{array}\right.
\end{equation}
and
\begin{equation}
\mu_k^{R}(N_i)= \left(\begin{array}{ccl}
-N_k + \sum_j [b_{jk}]_+ N_j & \qquad & i=k \\ \\
N_i& & \hbox{otherwise}
\end{array}\right.
\end{equation}
Note that these two result is the same rule if $\sum_j [b_{jk}]_+ N_j
=\sum_j [b_{kj}]_+ N_j$. For example, anomaly cancelation condition
of 4d $\cN=1$ theories of quiver type demands precisely this identity
for each and every node, and the familiar Seiberg duality map on
$N_i$ coincides with either of $\mu_{k}^{L,R}$.\footnote{ The mutation 
rule for the ranks can differ for different theories in various
dimensions. See Refs.\cite{Benini:2014mia,Benini:2011mf,Closset:2012eq,Xie:2013lya}
for 2d and 3d examples.}

There is another natural choice of mutation rule on $N_i$'s:
Mutate the total charge $\Gamma= \sum_i N_i\gamma_i$ as if it is one
of nodes in the quiver \cite{Manschot:2013dua}, i.e.,
\begin{equation}
\tilde\mu_k^{L}(\Gamma)= \left(\begin{array}{ccl}
-\Gamma & \qquad & \Gamma=n\gamma_k,\quad n\in {\mathbf Z}_+ \\ \\
\Gamma + [\langle \gamma_k, \Gamma\rangle ]_+ \gamma_k && \hbox{otherwise}
\end{array}\right.
\end{equation}
and
\begin{equation}
\tilde \mu_k^{R}(\Gamma)= \left(\begin{array}{ccl}
-\Gamma & \qquad & \Gamma=n\gamma_k,\quad n\in {\mathbf Z}_+  \\ \\
\Gamma + [\langle \Gamma,\gamma_k\rangle ]_+ \gamma_k && \hbox{otherwise}
\end{array}\right.
\end{equation}
where we introduced the notation $\tilde\mu$ to emphasize that
the shifts of $N_i$'s are different. Interestingly, their
action, when translated to that on $N_i$, boils down to a common
rule,
\begin{equation}\label{tmuN}
\tilde \mu_k^{L,R}(N_i)= \left(\begin{array}{ccl}
-N_k + {\rm min}\left(\sum_j [b_{jk}]_+ N_j,\sum_j [b_{kj}]_+ N_j\right)  & \qquad & i=k \\ \\
N_i& & \hbox{otherwise}
\end{array}\right.
\end{equation}
Otherwise $\tilde\mu$'s act on $\gamma$, $b$, $\zeta$, in the
same way as $\mu$'s. We will later see that this modified
mutation preserves the quiver invariant.

\section{$\Omega_{\rm Q}(\zeta)$ via Localization: Summary of HKY}

As a preliminary, we will review the HKY index formula
for the quiver quantum mechanics.
The equivariant Witten index of interest is
\begin{equation}
\Omega_{\rm Q}(\zeta)=\lim_{\beta\rightarrow \infty}
{\rm tr} \left[(-1)^{2J_3} {\bf y}^{2J_3+2I}e^{-\beta H(\zeta)}\right] \ ,
\end{equation}
where we fixed the usual sign ambiguity of the index by choosing
$(-1)^F=(-1)^{2J_3}$. When we do this we should take care to remove the
center of mass part of the low energy dynamics, which is to say, to remove
one overall $U(1)$ decoupled from the rest of the dynamics.

For GLSM with compact classical moduli space, the localization procedure
produces relatively compact finite integration over vector multiplet zero
modes. Denoting collectively by $u= \beta(\bar x_3+i \bar A_0)$ the
zero modes of Cartan part of the vector multiplet, the Witten index
for 1d ${\cal N}=4$ GLSM is  compactly expressed as a residue integral
of the following expression
\cite{HKY}\footnote{See also Refs.~\cite{Cordova,Hwang} for related discussions. }
\begin{eqnarray}
g(u)=\prod_A  g_{\rm gauge}^{(A)}(u)\prod_Ig_{\rm chiral}^{(I)}(u)\ ,
\end{eqnarray}
which comes from one-loop determinant of nonzero modes. For instance,
each gauge sector, labeled by $A$, contributes
$$
g_{\rm gauge}^{(A)}(u)= \left[
\left(\frac{1}{2\sinh[z/2]}\right)^{r_A} \cdot
\prod_{\alpha\in \Delta_A}
\frac{\sinh[\alpha(u)/2]}{\sinh[(\alpha(u)-z)/2]}\right] \ ,
$$
where $r_A$ is the rank of the gauge group, $e^{z/2}={\bf y}$, and $\alpha$'s
are root vectors. A chiral multiplet of charge $q_I$, with respect to
the Cartan, and $R$-charge $R_I$ gives
$$
g_{\rm chiral}^{(I)}(u)=(-1)\cdot\frac{\sinh[(q_I(u)
+(R_I/2-1)z+f_I\cdot a) /2]}{\sinh[(q_I(u)+R_Iz/2+f_I\cdot a)/2]}\ ,
$$
where $a$ collectively denotes flavor chemical potentials and $f_I$
the charges of the chiral multiplet under  flavor symmetries.

The space spanned by the Cartan zero modes $u$ is product of cylinders
$\left({\mathbb C}^*\right)^r$ where $r=\sum_A r_A$ is the total
rank. For quiver theories with the gauge group $\prod_A U(d_A)$,
we have $r+1=\sum_Ad_A $ since the overall $U(1)$ is decoupled.
This zero mode space is riddled with singular hypersurfaces defined by
poles of $g(u)$, such as $\alpha(u)=z$ and $q_I(u)+R_Iz/2+f_I\cdot a=0$, and the
Witten index is expressed as sum of iterated residues at co-dimension
$r$ singularities. The main technical difficulty is which of such poles contribute
and with what residue. Details of this was derived in HKY, to which
readers are forwarded, and here we will summarize the result.
The result is compactly expressed in terms of Jeffrey-Kirwan residue ~\cite{JK,SV} as
\begin{equation} \
\Omega_{\rm Q}= \frac{1}{|W|}\hbox{JK-Res}_\eta \left[g(u)d^ru\right] \ ,
\end{equation}
where $W$ is the Weyl group and $\eta$ is an arbitrary but generic vector
living in the vector space generated by charges $\{ Q \}=\{ \alpha \}\cup\{q\}$.
The above residue is a summation over all co-dimension
$r$ singularities in $\left({\mathbb C}^*\right)^r$ that can be defined as
the intersection of hyperplanes via a collection of charges $\{Q_i\}$.
A singularity where poles
due to $r$ charges $\{Q_{i_p}\}$ collide will contribute a term, computed
via the JK-residue formula,
\begin{eqnarray}\label{JK}
\hbox{JK-Res}_{\eta:\{Q_{i_p}\}}
\frac{d^r u}{(Q_{i_1}\cdot u)(Q_{i_2}\cdot u)\cdots(Q_{i_r}\cdot u)}
=\frac{1}{|{\rm Det} Q|} \ ,
\end{eqnarray}
if $\eta$ is a positive linear span of $\{Q_{i_p}\}$; otherwise,
JK residue is declared to be zero.
We implicitly allowed constant shift of
the pole location for notational convenience.

A couple of important points need to be clarified before we can
make actual use of this formula. Recall that
quiver quantum mechanics, and more generally GLSM quantum
mechanics,  undergo wall-crossing under
continuous change of FI constants. $\Omega_{\rm Q}$ is therefore
a piece-wise constant function of $\zeta$. This  aspect is hidden
in the fact that, in  $\left({\mathbb C}^*\right)^r$ spanned
by $e^u$'s, there are additional poles located at ${\rm Re}\,
u= \pm \infty$ or $\bar x_3=\pm \infty$. Subtlety in dealing
with this additional singularity results in the wall-crossing phenomena.
Here we will be content with giving a prescription. The simplest
way to achieve this is to assign a charge $Q_\infty\equiv -\zeta$
to this asymptotic region\footnote{HKY introduced $Q_\infty$
in a more limited sense when defining an integrand in the
intermediate step, rather than as an effective charge entering
the JK test. For this limited use, $Q_\infty=\zeta$  chosen
there works equally well such that with $\eta=\zeta$, the
asymptotic contribution vanish. }
and reject or accept the pole at such
places using the same JK residue test with $Q_\infty$ as one of $Q_i$'s,
\begin{equation}
\{ Q'\}=\{ \alpha \}\cup\{q\} \cup \{Q_\infty=-\zeta\}\ .
\end{equation}
We need to remind ourselves that the hyperplane associated with
$Q_\infty$ is the asymptotic boundary of $\left({\mathbb C}^*\right)^r$.
Once this is understood, a natural choice of $\eta$ emerges;
If one takes $\eta=\zeta$, where $\zeta$ is now embedded
into the charge vector space, the JK positivity test always
rejects $Q_\infty\equiv -\zeta$, meaning the poles located
at the asymptotic region of $\left({\mathbb C}^*\right)^r$
can be made to be irrelevant for the Witten index.

However, sometimes this choice is not available because $\zeta $ is
not generic, i.e., is spanned by less than $r$ charges. In such cases,
we may try to shift $\eta$ slightly away from $\zeta$ but still the
asymptotic poles do not contribute. To see how this can be achieved,
consider a small deformation $\delta $ such that $\eta=\zeta+\delta$.
We wish to see for what choices of $\delta$ the additional charge
$Q_\infty$ cannot pass the JK positivity test. Suppose it does for
some $\delta$ and some collection $\{Q_\infty, Q_{i_2},\dots,Q_{i_r}\}$, i.e.,
$$\zeta+\delta=b_1 Q_\infty +\sum_{p=2}^{r}b_i Q_{i_p}$$
with $b_{1,2,\dots,r}>0$. This implies
$$\zeta+\frac{1}{1+b_1}\delta=\sum_{p=2}^{r}\frac{b_p}{1+b_1}Q_{i_p}\ ,$$
so that  a straight line between $\zeta$ and $\eta$ in the
charge vector space encounters a wall spanned by a collection of
$r-1$ charges. A rank $r$ charge vector spaces can be divided into
chambers by walls which are positive spans of $r-1$ physical
charges, which is not to be confused with the physically distinct
chambers in the wall-crossing sense defined on the FI constant
space. We conclude  that as long as $\eta$ lives in the same
chamber as $\zeta$ in the charge vector space, the asymptotic
pole never enters the JK residue formula.
We will be making such choices in all of following computations,
and deal only with the hyperplanes associated with the
physical charges $\{ Q \}=\{ \alpha \}\cup\{q\}$.

This naive procedure encounters much difficulties when, at
a contributing pole, more than $r$ such hyperplanes
meet. For these so-called degenerate cases, the residue computation
depends on the order of integration
 and the contribution from such a
point consists of several such iterated residues. This reflects
the fact that the middle homology of the Cartan zero mode space
at such a singularity is no  longer generated by a single cycle
and the integral required is a sum of integrals over
several such. A couple of constructive procedures are available
to deal with such cases, details of which will not
be discussed here, as they are available elsewhere \cite{BV,SV}.
In this note we follow a constructive procedure of Ref.~\cite{SV},
as described by Benini et. al. \cite{Benini:2013xpa}.

Finally, we wish to point out that this derivation is performed
with finite $\beta$ rather than by taking $\beta\rightarrow\infty$
limit, and can thus potentially fail to capture the true index. This
is remedied by taking large $\zeta$ limit while maintaining
the chamber \cite{HKY}, which suffices for theories with compact
classical moduli space or otherwise by adding enough chemical
potential to lift flat directions. There are examples of GLSM
for which these remedies are not enough to lift asymptotic
flat directions, but this goes beyond the scope of this note.

\section{Wall-Crossing and Quiver Invariants}

Wall-crossing, which is unique to 1d theories, is a
discontinuity of supersymmetric spectra in the $\zeta$ space.
The co-dimension-one ``walls" in the $\zeta$ space are
defined as $\sum_i n_i\zeta_i=0$, where we also have
$\sum_i (N_i-n_i)\zeta_i=0$. At such places, the phases
of central charges of $\Gamma_1=\sum_i n_i\gamma_i$
and $\Gamma_2= \sum_i (N_i-n_i)\gamma_i$ coincide precisely, and
if both $\Gamma_{1,2}$ exist as BPS states, wall-crossing occurs
such that degeneracy of $\Gamma=\Gamma_1+\Gamma_2$ can change
suddenly across the wall or at the wall. At the level of equivariant index,
this can be phrased as piece-wise constant behavior of $\Omega_{\rm Q}(\zeta)$
in the space of $\zeta$.

One intuitive way to understand this discontinuity is
to consider the so-called ``Coulomb" description of the quiver
theory, where only the Cartan part of $U(N_i)$'s are kept
and all other degrees of freedom are integrated out. Naively,
this picture is valid when all $\zeta$'s are small relative to the
scale of 1d gauge couplings. One ends up with a
collection of $\sum N_i$ charged particles in ${\mathbf R}^3$
space where the vector multiplet scalars live in, and the
ground states look like multi-center bound states where
individual centers of charge $\gamma_i$'s are balanced against
one another by combination of attractive Coulomb-like potential
and repulsive ``angular momentum" barrier.

The mutual distances of these constituent particles are set
by inverses of $\zeta_i$'s, and the wall-crossing discontinuity happens
as one or more charged particles, say of total charge $\sum_in_i\gamma_i$,
move off to infinity of ${\mathbf R}^3$, relative to the others
when  $\sum_i n_i\zeta_i$ vanishes. The discontinuity of index
occurs because such a state fails square-normalizability.
Very general state counting in this picture
has been carried out in recent years, which we will denote
collectively as
\begin{equation}
\Omega_{\rm Q}^{\rm Coulomb}(\zeta) = {\rm Tr}\,(-1)^{2J_3} {\bf y}^{2I+2J_3} e^{-\beta H_{\rm Q}^{\rm Coulomb}(\zeta)}
\end{equation}
and which has been compared successfully, for quivers
without oriented loops and thus without superpotential,
against various mathematical results such as Reineke's
results \cite{Reineke} and results deduced from Kontsevich-Soibelman \cite{KS}
wall-crossing algebra.

However, it turns out that this ``Coulomb" picture can
miss a huge set of ground states when the quiver admits
superpotentials. These additional ground states remain
centered and compact near the origin even while $\sum_i n_i\zeta_i \rightarrow 0$, and
thus easily survive wall-crossing catastrophe. The quiver
invariant can be defined as those states that survive at
all the ``walls," and they continue to exist as square-normalizable
wavefunctions at the  intersection of all marginal
stability wall. In 1d GLSM, the latter corresponds
to the origin of FI constant space, $\zeta_i=0$ for all $i$.
Counting the index at such a place, one is naturally lead to
the definition of the quiver invariant
\begin{equation}
\Omega_{\rm Q}\biggl\vert_{\rm Inv}=\lim_{\beta\rightarrow\infty} {\rm Tr}_{L^2}
\,(-1)^{2J_3} {\bf y}^{2I+2J_3} e^{-\beta H_{\rm Q}(\zeta=0)}\ .
\end{equation}
Although we will be mostly concerned with quiver theories in
this note, it is clear that the same definition can be
extended to  other 1d GLSM theories, defining GLSM invariants
in a similar manner.
Note that we took care to impose $L^2$ condition on wavefunctions,
as $\zeta=0$ generically results in asymptotic runaway directions
along the vector multiplet scalars. Otherwise, the quantity would
be either ill-defined or could give misleading answers. This also
tells us direct evaluation will be pretty difficult.

For Abelian cyclic quivers, this split between Coulombic
multi-center states and the wall-crossing-safe quiver invariants
is clean and has been understood rigorously. For example,
let us take triangle quiver with $(b_{23},b_{31},b_{12})=(4,5,6)$.
The quiver admits three different chambers, where the Hodge
diamonds turn out to be
\begin{equation}
\begin{array}{ccccccccccc}
&&&&&1&&&&& \\
&&&&0&&0&&&& \\
&&&0&&2&&0&&&\\
&&0&&0&&0&&0&&\\
&0&&0&&3&&0&&0&\\
0&&0&&26&&26&&0&&0\\
&0&&0&&3&&0&&0&\\
&&0&&0&&0&&0&&\\
&&&0&&2&&0&&&\\
&&&&0&&0&&&& \\
&&&&&1&&&&& \\
\end{array} \ ,
\qquad
\begin{array}{ccccccc}
&&&1&&& \\
&&0&&0&& \\
&0&&2&&0&\\
0&&26&&26&&0\\
&0&&2&&0&\\
&&0&&0&& \\
&&&1&&& \\
\end{array} \ , \qquad
\begin{array}{ccc}
&1& \\
26&&26 \ .\\
&1& \\
\end{array}\nonumber
\end{equation}
For this simple class, the relevant geometry is entirely
toric or a complete intersection therein, so the
cohomology is easy to compute.

One important observation,  emerged from study of
these cyclic Abelian quivers, is that states counted by
$\Omega\vert_{\rm Inv}$ are always $SU(2)_R$ singlet
but can be charged under $U(1)_R$, while those counted
by $\Omega^{\rm Coulomb}$ are neutral under $U(1)_R$ and
typically in $SU(2)_R$ multiplet. In the low energy
nonlinear sigma model limit, or equivalently in the
so-called ``Higgs" description, $2J_3$ and $2I$ labels
the vertical and the horizontal directions of the Hodge
diamond. For the above example, $26+26 $ states in the
horizontal middle belong to $\Omega\vert_{\rm Inv}$.
It turns out that these features of Coulombic states
and wall-crossing safe states being, respectively,
vertical and horizontal  middle cohomology elements
are completely general.

Any wavefunction of multi-center nature will loose its
square-normalizability upon $\zeta=0$, among which are states
counted by $\Omega_{\rm Q}^{\rm Coulomb}(\zeta)$. However, for more
general quivers, there are also hybrid type of multi-center
states where, among the constituents ``particles," one finds
quiver invariants of subquivers. Therefore, if one is to
study supersymmetric ground states in such multi-center
viewpoint, one must count many different kinds of multi-center
bound states with both elementary constituents and those
from quiver invariants of subquivers.

This physically compelling idea has been consolidated
into a partition sum identity as follows \cite{Manschot:2013sya,Manschot:2014fua},
\footnote{Because this formula originates from a form of Abelianization
routine which is natural in Coulombic construction of vacuum states,
the precise formula  involves various combinatoric factors due to
Weyl projections and needs to be phrased via the rational version
of the index, $\bar \Omega$; here, we refer readers to existing literatures
\cite{Manschot:2010qz,Manschot:2011xc,Kim:2011sc}}.
\begin{equation}\label{MPS}
\Omega_{\rm Q}(\zeta)\sim \sum_{{\rm Q}=\oplus_p {\rm Q}_p} \Omega_{{\rm Q}/\{{\rm Q}_p\}}^{\rm Coulomb}(\{\zeta_p\}))
\times \left(\prod_p\Omega_{{\rm Q}_p}\biggl\vert_{\rm Inv}\right) \ .
\end{equation}
The right hand side requires further explanation.\footnote{In
Refs.~\cite{Manschot:2013sya,Manschot:2014fua}, the counterpart of
$\Omega\vert_{\rm Inv}$ is denoted as $\Omega_S$, where S stands for
single-center states.  Furthermore, the authors proposed this
expansion formulae for Poincar\'e polynomials rather than for
indices, so that $\Omega_S$ of theirs is actually an integer
rather than Laurent polynomials of ${\bf y}$. However, the same
expansion formula should work for indices provided that pure
Coulombic wavefunctions have vanishing $R$-charges and that all
states counted by the quiver invariants are $SU(2)_R$ singlets.}
The sum is over all possible partition of the quiver, which is
to say all possible partitions of the ranks, $N_i=\sum_p N_i^{(p)}$,
with nonnegative integers $N_i^{(p)}$'s.
Each such partition defines a set of subquivers ${\rm Q}_p$ with
ranks $N_i^{(p)}$. The adjacency matrix $b$ and FI constants
$\zeta$'s of ${\rm Q}_p$ are the same as those of $\rm Q$.
The quiver denoted as ${\rm Q}/\{{\rm Q}_p\}$ is an induced quiver where
each of subquivers ${\rm Q}_p$ is treated as if it is a single node
of charge $\Gamma_p=\sum_i N_i^{(p)}\gamma_i$. The induced
adjacency matrix and the induced FI constants of ${\rm Q}/\{{\rm Q}_p\}$
are determined naturally, e.g., $b_{pq}=\langle \Gamma_p,\Gamma_q\rangle$
and $\zeta_p= \sum_i N_i^{(p)}\zeta_i$ etc.

The simplest
nontrivial example is again the Abelian cyclic quivers,
which motivated the above partition
sum to begin with. In the latter class, the summation
consists of only two terms,
\begin{equation}
\Omega_{\rm Q}(\zeta)=1\times \Omega_{\rm Q}\biggl\vert_{\rm Inv}
+\;\;\Omega_{\rm Q}^{\rm Coulomb}(\{\zeta_i\}))\times\left(\prod_i 1\right)\ .
\end{equation}
The first term corresponds to ${\rm Q}={\rm Q}_1$, i.e. $N_i^{(1)}=N_i=1$,
such that ${\rm Q}/\{{\rm Q}_p\}$ is the trivial single node Abelian quiver.
The second corresponds to ${\rm Q}=\oplus_p {\rm Q}_p$ with
$N_i^{(p)}=\delta_i^p$, so that ${\rm Q}/\{{\rm Q}_p\}$ is ${\rm Q}$ itself.
Finally ``1" factors are associated with the elementary
and free $U(1)$ quiver, which signals the underlying object,
4d quantum state in half-hypermultiplet or the rigidly
wrapped D-brane. All other subquivers are tree-like
with vanishing $\Omega\vert_{\rm Inv}$ and are thus
absent in the sum. In this class of quivers, states
counted by the first spans horizontal middle of the
Hodge-decomposed cohomology which remains robust under
any of the wall-crossing, while the second spans the
vertical middle and changes chamber by chamber.
Abelian cyclic quivers are a little special,
as states counted by $\Omega_{\rm Q}^{\rm Coulomb}$ can be given
special geometric meaning \cite{Lee:2012naa}, via Lefschetz hyperplane theorem,
but generalization of this to general quiver is not known.

This partition sum actually goes further than a mere
reproduction of true index via multi-center viewpoint.
Eq.~(\ref{MPS}), whose idea should extend to the
supersymmetric Hilbert space itself, means that one
can reconstruct the entire Hodge diamonds, or the entire
supersymmetric spectra for any given quiver ${\rm Q}$.
For general quivers, especially
those involving $N_i>1$ for some $i$, the cohomology
computation is mathematically very challenging. The possibility
of a computationally straightforward determination of
cohomologies of entire class of quiver varieties is quite
remarkable, to say the least.

To illustrate this, take a non-Abelian triangle quiver of
ranks $(1,1,3)$ and  the adjacency matrix with
$(b_{23},b_{31},b_{12})=(3,5,10)$. Let us denote their
indices as $\Omega^{1,1,3}_{3,5,10}$. The relevant quiver
invariants are $\Omega^{1,1,N}_{3,5,10}\vert_{\rm Inv}$ for
$N=1,2,3$, as no other subquiver can have a quiver invariant.
It turns out that
\begin{eqnarray}
\Omega^{1,1,1}_{3,5,10}
\biggl\vert_{\rm Inv}&=&0\ ,\cr
\Omega^{1,1,2}_{3,5,10}\biggl\vert_{\rm Inv}&=&6/{\bf y}+6{\bf y} \ ,\cr
\Omega^{1,1,3}_{3,5,10}\biggl\vert_{\rm Inv}&=&0\ .
\end{eqnarray}
Thus,  there are only
two nontrivial terms in the partition sum;
\begin{equation}
\Omega^{1,1,3}_{3,5,10}=  \left(\Omega^{1,1}_{2}\right)^{\rm Coulomb} \times \Omega^{1,1,2}_{3,5,10}\biggl\vert_{\rm Inv}\times 1
+\;\;\left(\Omega^{1,1,3}_{3,5,10}\right)^{\rm Coulomb}\times 1^5 \ .
\end{equation}
One is the maximal partition, for which ${\rm Q}/\{{\rm Q}_p\}={\rm Q}$
itself.  The other is $(1,1, 3)=(1,1,2)\oplus(0,0,1)$ for which
${\rm Q}/\{{\rm Q}_p\}$ is a two-node Abelian quiver with the intersection number
$\langle \gamma_3 , \gamma_1+\gamma_2+2\gamma_3\rangle =5-3=2$.

For example, in the chamber of the maximal
moduli space dimensions, the Witten index is \cite{HKY}
\begin{equation}
\Omega^{1,1,3}_{3,5,10}
= 1/{\bf y}^6 + 2/{\bf y}^4 -2/{\bf y}^2 -7  -2 {\bf y}^2 + 2 {\bf y}^4 + {\bf y}^6\ ,
\end{equation}
while the relevant Coulomb indices are
\begin{eqnarray}
\left(\Omega^{1,1}_{2}\right)^{\rm Coulomb}&=&-1/{\bf y}-{\bf y}\ , \cr\cr
\left(\Omega^{1,1,3}_{3,5,10}\right)^{\rm Coulomb}&
=& 1/{\bf y}^6 + 2/{\bf y}^4 + 4/{\bf y}^2 + 5  + 4 {\bf y}^2 + 2 {\bf y}^4 + {\bf y}^6
\ .
\end{eqnarray}
{}From these, we can reconstruct the Hodge diamond
\begin{equation}
\begin{array}{ccccccccccccc}&&&&&&&&&&&& \\
&&&&&&1&&&&&& \\
&&&&&0&&0&&&&& \\
&&&&0&&2&&0&&&&\\
&&&0&&0&&0&&0&&&\\
&&0&&0&&4&&0&&0&&\\
&0&&0&&6&&6&&0&&0&\\
0&&0&&0&&5&&0&&0&&0 \\
&0&&0&&6&&6&&0&&0&\\
&&0&&0&&4&&0&&0&&\\
&&&0&&0&&0&&0&&&\\
&&&&0&&2&&0&&&&\\
&&&&&0&&0&&&&& \\
&&&&&&1&&&&&& \\&&&&&&&&&&&& \\
\end{array}
\end{equation}
of this chamber. Other chambers can be treated similarly.

Currently, however, we do not know of any direct and practical
computational method for the quiver invariants. In the
above example, we actually computed $\Omega_{\rm Q}$'s and
$\Omega_{\rm Q}^{\rm Coulomb}$'s first, and then inferred
$\Omega_{\rm Q}\vert_{\rm Inv}$ inductively. Note that the
Witten index by itself cannot give us the full cohomology
information. Construction of
the Hodge diamond comes as a bonus along the process.
Absence of a direct computational tool for $\Omega_{\rm Q}\vert_{\rm Inv}$,
despite its very elegant and robust nature, is unsatisfactory.
One purpose of this note is to consider how mutation might help
us in determining $\Omega_{\rm Q}$'s and $\Omega_{\rm Q}\vert_{\rm Inv}$
by mapping high-rank quivers to lower-rank ones.

\section{Mutation $\mu_k$ on Witten Index $\Omega_{\rm Q}(\zeta)$}

\subsection{Mutations and Chambers}

Mutation is a Seiberg-duality on quiver gauge theories,
and expected to preserve physics. Depending on dimensions,
it works slight differently. In 4d, a node has the same number
of fundamental and anti-fundamental fields, due to gauge anomaly
cancelation, so $\mu^L=\mu^R$. In 2d, the equality between
the incoming arrows and the outgoing arrows is connected
to Calabi-Yau condition and no longer necessary for consistency;
in principle $\mu^L$ and $\mu^R$ can induce two different dualities,
although Benini et.al.\cite{Benini:2014mia} argued that, for each
mutation step at node $k$, one must choose one of $\mu^{L,R}_k$
for which $\mu_k(N_k)$ is the larger. In 1d, even the choice of
mutation node is restricted such that, given a point in $\zeta$
space, one could mutate at most two nodes, one by $\mu^L$
and the other by $\mu^R$.

This happens, for ${\cal N}=4$ quiver quantum mechanics,
because of wall-crossing phenomena. Given $\zeta$'s fixed,
the mutation is not
allowed for all nodes. Physically clean criteria, applied
to rank-one Seiberg-Witten theories, were given by Alim et. al.
\cite{Alim:2011ae,Alim:2011kw},
who argued that mutation should be thought of as change of
basis charges. Here the basis means that the rest of BPS
charges can be built as a sum over the basis with
non-negative integer coefficients. What classifies
a charge as BPS instead of anti-BPS is an arbitrary convention,
so by rotating the relevant ``upper-half-plane" in the central
charge plane, one is sometimes forced to give up a basis
element $\gamma_k$, in favor of $-\gamma_k$. This can
affect the rest of basis as well, and $\mu^{L,R}_k$ we
introduced earlier were proposed to be the correct
transformation of basis under such rotation of ``upper-half-plane."

Note that this rotation of upper-half-plane mutates
one charge at a time, and the choice is not random. The basis
element to be mutated has to be the closest to the other
lower-half-plane, either along the right-side of the half planes
or along the left-side. Thus, one can anticipate
that the left(right) mutation will leave physics invariant only when
acting on very specific charge. In terms of quiver theories,
this translates to inequalities among FI constants; Along the real
axis of FI constants, we are allowed either to mutate-left the
left-most node or to mutate-right the right-most node.

\begin{figure}[h]
\centering
\includegraphics[width=1\textwidth]{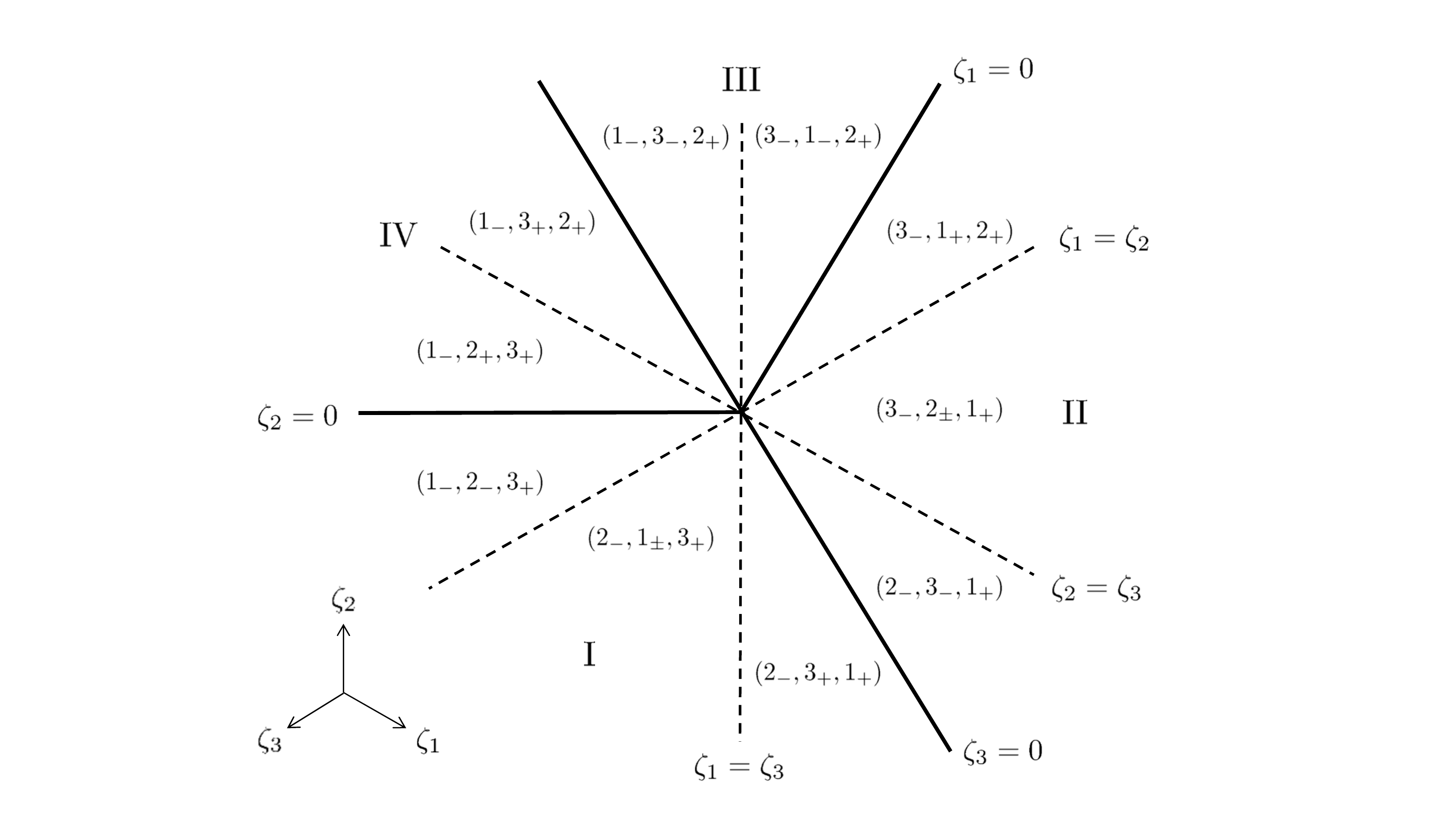}
\caption{\small  Four physical chambers of $(1,1,N)$ triangle quivers,
divided by solid lines. These are further divided into ten
sub-chambers by relative ordering of the three FI constants;
for example, $(2_-,3_-,1_+)$ means $\zeta_2<\zeta_3<0<\zeta_1$.
The arrows in the  lower-left corner are normal to the respective constant
$\zeta$ lines.}
\label{sub}
\end{figure}

\begin{figure}[h]
\centering
\includegraphics[width=1\textwidth]{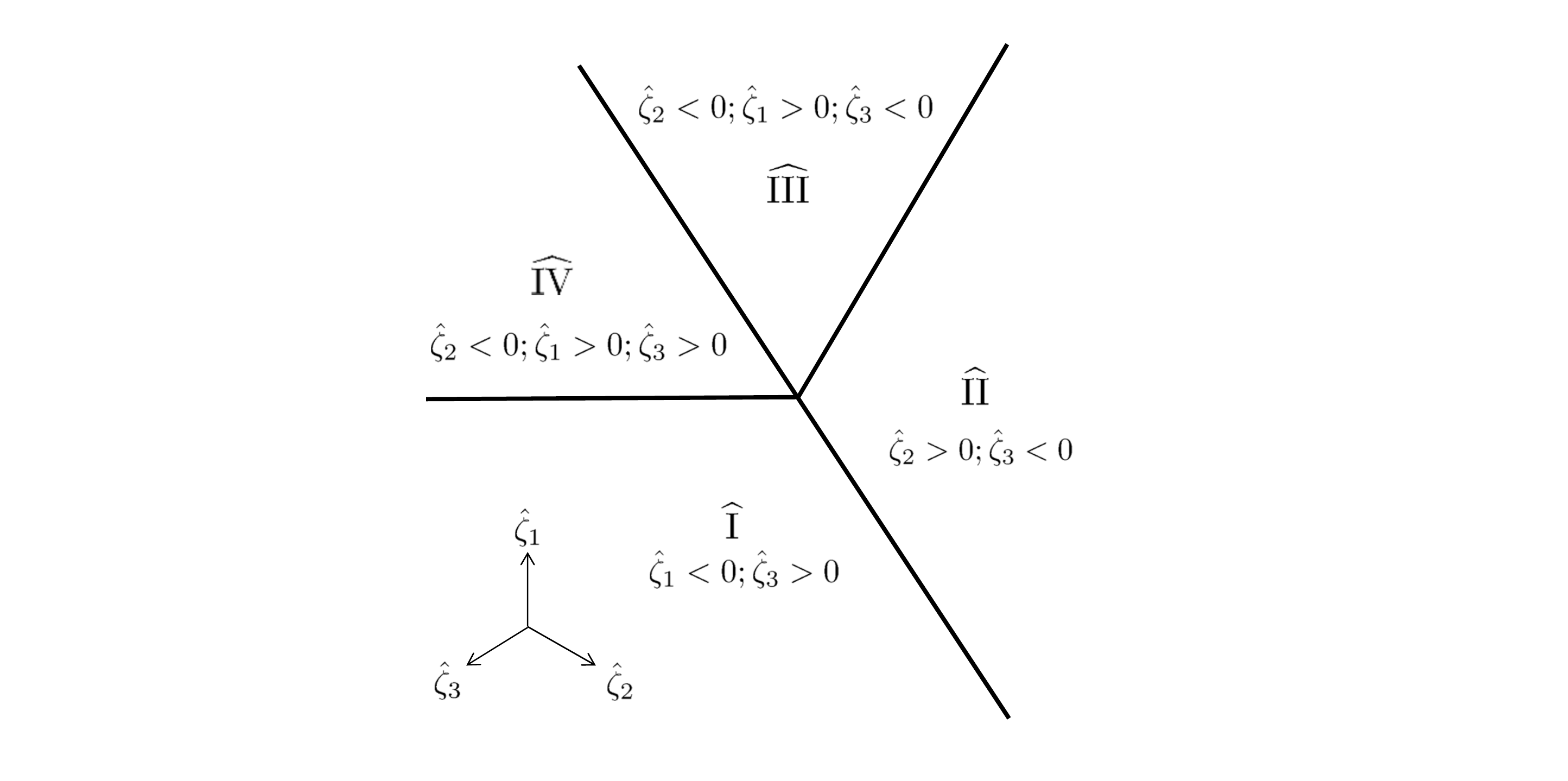}
\caption{\small  Mutating on node 3 of $(1,1,N)$ quiver brings us
back to another $(1,1,\widehat{N})$ quiver. Because the mutation
flips arrow orientations, the roles of $\hat\zeta_1$ and
$\hat\zeta_2$ are exchanged relative to those of $\zeta_{1,2}$.
The left mutation, allowed in three sub-chambers of figure 2
with most negative $\zeta_3$, maps
indices of chambers II and III, respectively, to those of
chambers $\widehat{\rm  IV}$ and $\widehat{\rm I}$. Similarly,
the right mutation, allowed in three sub-chambers of figure 2
with most positive $\zeta_3$,
maps indices of chambers I and IV, respectively,
to those of chambers $\widehat{\rm III}$ and $\widehat{\rm II}$.}
\label{afterM}
\end{figure}

One immediate question to be asked here is what happens if $\mu_k(N_k)$
happen to be negative for some $k$. The mutation is ill-defined because
a negative rank appears in the node $k$ of $\widehat{Q}$. Does this simply
mean that the mutation map become unavailable? Or could there be still
additional information about the original quiver?  Let us observe
that the index of the original quiver vanishes whenever
$\gamma_k$ is the left-most (right-most) and $\mu_k^{L}(N_k)<0$
($\mu_k^{R}(N_k)<0$), which follows from the $D$-term condition at node $k$,
\begin{equation}
X_k X_k^\dagger -Y_k^\dagger Y_k=\zeta_k I_{N_k\times N_k}\ ,
\end{equation}
where $X_k$ is a rectangular complex matrix of $N_k\times (\sum_j[b_{jk}]_+N_j)$
type and collectively denotes all chiral multiplets associated
with incoming arrows. $Y_k$ is of type $(\sum_j[b_{kj}]_+N_j) \times N_k$
and represents the collection of all outgoing arrows.
When $\zeta_k$ is
negative (positive), the right hand side is of rank $N_k$ with all
negative (positive) and equal eigenvalues, and this $D$-term equation
can be solved only if $Y$ ($X$) is of rank $N_k$ also.
When $\zeta_k$ is
the left-most (right-most) and thus necessarily negative (positive),
this condition for non-empty moduli space translates,
upon the respective mutation, to $\mu_k^{L}(N_k)\ge 0$
($\mu_k^{R}(N_k)\ge 0$). Therefore we conclude that whenever
a formally valid mutation results in a negative rank of the mutated
node, the original quiver must have been in a physically empty chamber
with a vanishing Witten index.  In this sense, it suffices to consider
the original quivers and the chambers thereof such that allowed
mutation results in $\mu_k(N_k)\ge 0$, to which cases we will restrict
ourselves.

With the index counting enabled by HKY's general formula, we wish
to test this mutation idea explicitly by applying to a simplest
class of triangle quivers. We will perform numerical test
as well as illustrate how HKY formula itself exhibits invariance
under such mutations. The latter may be generalized to a larger
class of quivers, establishing the mutation invariance rigorously
at the level of index theorem.

\subsection{A Numerical Check and A Subtlety}

\begin{figure}[h]
\centering
\includegraphics[width=0.8\textwidth]{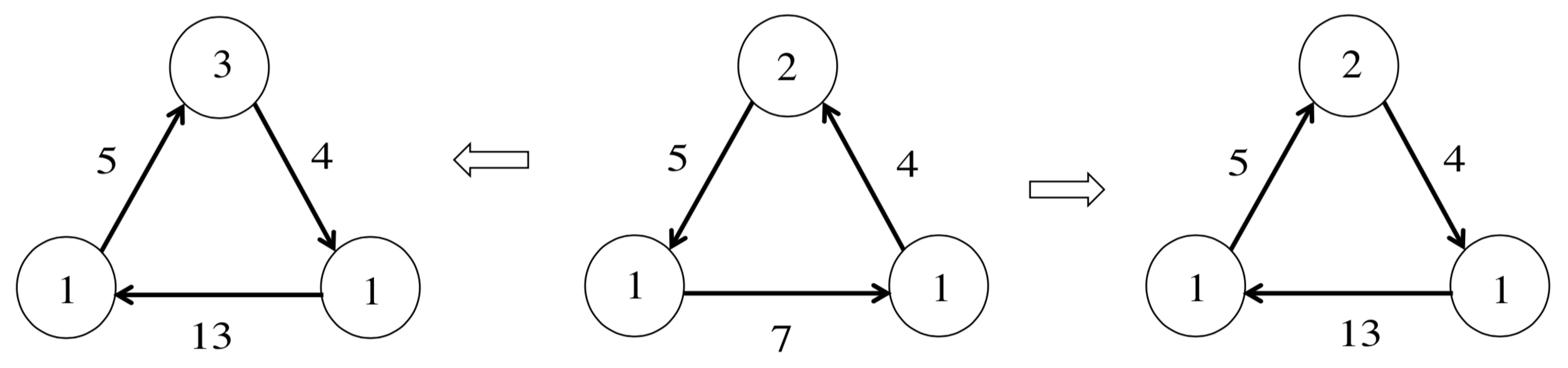}
\caption{\small  An explicit example of mutation. Witten
indices are computed for all four chambers for each of the three quivers,
showing that mutation selection rule is necessary.}
\label{example}
\end{figure}

Before we plunge into more analytical demonstration in next subsection,
let us briefly check the validity of the mutation invariance with
a particular example of triangle quivers with ranks $(1,1,2)$ and the
intersection numbers $(4,5,7)$ of figure~\ref{example}. This will
serve to check the aforementioned assertion, regarding invariance of Witten
indices of particular chambers as well as non-preservation of Witten
indices of ``wrong" chambers.
Indices of the original quiver were computed in Ref.~\cite{HKY},
\begin{eqnarray}
\Omega({{\rm I}})&=&50\ ,\label{745}\cr\cr
\Omega({{\rm II}})&=& {1}/{{\bf y}^4}+
{2}/{{\bf y}^2}+87+2{\bf y}^2+{\bf y}^4\ ,\cr\cr
\Omega({{\rm III}})&=&{1}/{{\bf y}^6}+
{2}/{{\bf y}^4}+{4}/{{\bf y}^2}+89
+4{\bf y}^2+2{\bf y}^4+{\bf y}^6\ ,\cr\cr
\Omega({{\rm IV}})&=&{1}/{{\bf y}^6}+
{2}/{{\bf y}^4}+{4}/{{\bf y}^2}+54+4{\bf y}^2+2{\bf y}^4+{\bf y}^6\ .
\end{eqnarray}

Under the left mutation, we find  a quiver with ranks $(1,1,3)$ and
intersection numbers $(-4,-5,-13)$ with indices,
\begin{eqnarray}
\Omega({\widehat{\rm I}})&=&{1}/{{\bf y}^6}+
{2}/{{\bf y}^4}+{4}/{{\bf y}^2}+89
+4{\bf y}^2+2{\bf y}^4+{\bf y}^6 \ ,\cr\cr
\Omega({\widehat{\rm II}})&=&35\ ,\cr\cr
\Omega({\widehat{\rm III}})&=&{1}/{{\bf y}^4}+
{2}/{{\bf y}^2}+37+2{\bf y}^2+{\bf y}^4\ ,\cr\cr
\Omega({\widehat{\rm IV}})&=&{1}/{{\bf y}^4}+
{2}/{{\bf y}^2}+87+2{\bf y}^2+{\bf y}^4\ .
\end{eqnarray}
Note that $\Omega({\small\rm II})=\Omega({\small\widehat{\rm IV}})$ and
$\Omega({\small\rm III})=\Omega({\small\widehat{\rm I}})$, as anticipated.
Under the right mutation, we find a quiver with ranks $(1,1,2)$ and
intersection numbers $(-4,-5,-13)$ with indices,
\begin{eqnarray}
\Omega({\small\widehat{\rm I}})&=& {1}/{{\bf y}^{10}}+{2}/{{\bf y}^8}+{4}/{{\bf y}^6}+{6}/{{\bf y}^4}+{8}/{{\bf y}^2}\cr
&&\qquad\qquad +\ 58 + 8 {\bf y}^2+6 {\bf y}^4+4 {\bf y}^6+2 {\bf y}^8+{\bf y}^{10}\ ,\cr\cr
\Omega({\small\widehat{\rm II}})&=& {1}/{{\bf y}^6}+
{2}/{{\bf y}^4}+{4}/{{\bf y}^2}+54+4{\bf y}^2+2{\bf y}^4+{\bf y}^6,\cr\cr
\Omega({\small\widehat{\rm III}})&=&50\ ,\cr\cr
\Omega({\small\widehat{\rm IV}})&=&50\ .
\end{eqnarray}
We find that $\Omega({\small\rm I})=\Omega({\small\widehat{\rm III}})$ and
$\Omega({\small\rm IV})=\Omega({\small\widehat{\rm II}})$, again as anticipated.

Perhaps equally noteworthy is the fact that if one starts in
disallowed sub-chambers, where mutation on this node is not justified,
Witten indices before and after the mutation do not match. In fact,
even the dimension of the classical moduli spaces can differ before and
after mutation. This  example thus demonstrates that the selection
rules for the mutable node and choice of the  mutation orientation are
very much necessary.

Apart from checking the mutation invariance numerically, this
exercise gives a valuable hint on how to demonstrate
mutation invariance between a pair of $(1,1,N)$ type quivers.
For general quivers, classifying poles according to JK positivity
test poses a big combinatorial challenge. This is further aggravated
by the presence of degenerate poles where more than $r$ singular
hyperplanes collide. When such a degenerate pole passes JK positivity
test, the iterated residue becomes order-dependent and further
combinatorial task emerges.
Such technical issues, however, are much ameliorated when one can exclude
hyperplanes associated with vector multiplets from the analysis. This
not only reduces poles passing JK test drastically but also tends to
remove a lot of degenerate singularities.

For simple quivers, such as
primitive tree-like quivers, there is a reasonable argument why
JK-acceptable singularities involving the vector multiplet poles
must have a vanishing residue \cite{Benini:2013xpa}. This follows
from a counting of the net number of zeros against the
net number of poles. For other quivers,
such as our triangles with a loop, this argument does not extend
straightforwardly. For example, the pole due to the chirals
between nodes 1 and 2 can coincide with a vector multiplet
pole of node 3, such that the vanishing argument due to counting
of zeros and poles no longer works. Furthermore, singularities
of this type tend to fail the so-called projective property which
enables one to derive the residue formulae.

In the end, however, extensive numerical exercises with $(1,1,N)$
quivers lead us to believe that the vector multiplet poles need
not be considered at all for this class
of quivers.\footnote{Irrelevance of vector multiplet poles
is hardly a general statement. Counterexamples
include non-primitive Kronecker quivers as well as, more obviously,
$SU(n)$ gauged quantum mechanics without matter multiplets.
Note that these examples have flat Coulombic directions,
however, so the quantity computed by HKY is not true $L^2$ index.
Nevertheless, these examples shows that the existing prescription
rules cannot by themselves preclude vector multiplet poles.
 Establishing general criteria
on when we are allowed to ignore vector multiplet poles will
go a long way for our understanding of the Witten index of general GLSM.}
Most of the singularities involving vector multiplet poles and
also passing JK positivity test, can be seen to have
a vanishing residue straight-forwardly. The main issue is how to deal with those
non-projective singularities. We have regulated these by
shifting the coordinates to split them artificially to
projective ones, evaluate the residues, and ``unshift." The
reduced projective singularities give a vanishing residue,
again due to the vector multiplet poles being canceled by
chiral zeros, and we are back to the statement that vector
multiplet poles need not be considered. This simplifies
the problem enormously since for each chamber there is exactly
one iterated residue integral that contributes to the index.
Establishing duality between a mutation pair of $(1,1,N)$ quivers
amounts to showing these two residues agree with each other
regardless of the intersection numbers $(b, c, a)$ and $(-b,-c, a-bc)$,
which we will show in the next subsection.

\subsection{Mutation Invariance of Witten Index}

\begin{figure}[t]
\centering
\includegraphics[width=0.9\textwidth]{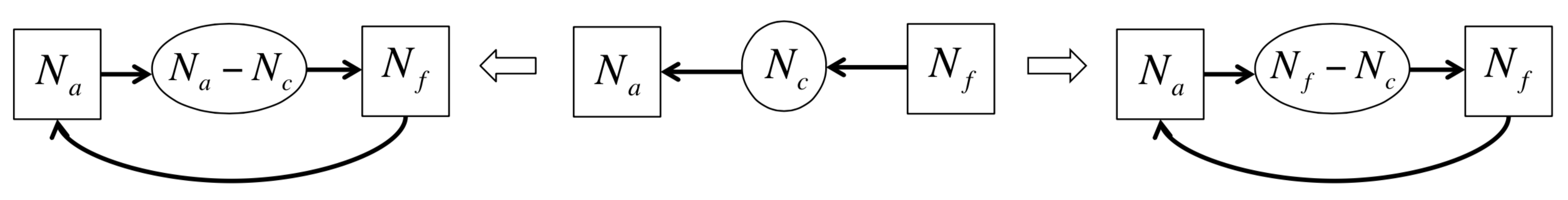}
\caption{\small  Seiberg-like dualities for 1d SQCD. There exist two different
duality maps depending on sign of the FI parameter.}
\label{SQCD}
\end{figure}

The prototype of mutation invariance for 1d GLSM can be
found in SQCD-like theories with a single $U(N_c)$
gauge group coupled to $N_f$ and $N_a$ number of fundamental
and anti-fundamental chirals, as drawn in the middle of
Figure \ref{SQCD}. The one-loop determinant of this theory
is
\begin{eqnarray}\nonumber
g(u,z)&=&\left(\frac{1}{2\sinh[z/2]}\right)^{N_c}
\prod_{i\neq j}\frac{\sinh[(u_i-u_j)/2]}{\sinh[(u_i-u_j-z)/2]}\cr
&&\times\prod_{i=1}^{N_c}\prod_{\alpha=1}^{N_f}
\frac{-\sinh[(u_i-a_\alpha+R_fz/2-z)/2]}{\sinh[(u_i-a_\alpha+R_fz/2)/2]}
\\
&&
\times\prod_{i=1}^{N_c}\prod_{\beta=1}^{N_a}\frac{-\sinh[(-u_i+b_\beta+R_az/2-z)/2]}{\sinh[(-u_i+b_\beta+R_az/2)/2]}\ ,
\end{eqnarray}
where $a_\alpha$'s and $b_\beta$'s are flavor fugacities.
Although $R$-charges in this simple theory are ambiguous
due to possible mixing with flavor and gauge charges, we keep
them explicitly as we wish to embed the theory to a larger
theory later on.

This theory has two chambers distinguished by sign of $\zeta$.
The index at $\zeta>0$ can be evaluated by sum over all
possible configurations of picking up $N_c$ poles in the
fundamentals,
\begin{eqnarray}\nonumber
\text{JK-res}_{\zeta>0}~g(u,z)&=&
\sum_{A \in C(N_f,N_c)}\prod_{\substack{i \in A \\ j \in A'}}
\frac{-\sinh[(a_i-a_j-z)/2]}{\sinh[(a_i-a_j)/2]}\\
&&\times\prod_{i\in A}\prod_{\beta=1}^{N_a}
\frac{-\sinh[(-a_i+b_\beta+(R_f+R_a)z/2-z)/2]}{\sinh[(-a_i+b_\beta+(R_f+R_a)z/2)/2]}\ ,
\end{eqnarray}
where summation is taken over $N_f \choose N_c$ choices of a set $A$
which choose $N_c$ fugacities of fundamentals. We also denote $A'$ by its complement.
Note that we can rewrite this expression as
\begin{eqnarray}
&&\sum_{A' \in C(N_f,N_f-N_c)}\prod_{\substack{i \in A \\ j \in A'}}
\frac{-\sinh[(-a_j+a_i-z)/2]}{\sinh[(-a_j+a_i)/2]}\cr
&&\qquad\qquad\times\:\:
\prod_{j\in A'}\prod_{\beta=1}^{N_a}
\frac{-\sinh[(a_j-b_\beta-(R_a+R_f)z/2)/2]}{\sinh[(a_j-b_\beta+(2-R_a-R_f)z/2)/2]}\cr
\label{dualf}
&&\qquad\qquad\times
\prod_{\alpha\in A\cup A'}\prod_{\beta=1}^{N_a}
\frac{-\sinh[(-a_\alpha+b_\beta+(R_a+R_f)z/2-z)/2]}{\sinh[(-a_\alpha+b_\beta
+(R_a+R_f)z/2)/2]}\ ,
\end{eqnarray}
which is nothing but the index of the theory with $U(N_f-N_c)$ gauge group
with same number of (anti-)fundamentals together with $N_fN_a$ mesons by
superpotential $W={\rm Tr}\,\bar\Phi M\Phi$.
Especially, the theory is mapped to the chamber
with $\zeta'<0$ of the dual theory.
Note that the $R$-charges of the dual quiver is shifted to $(R_f',R_a',R_a+R_f)$
where $R_f'+R_a'=2-R_f-R_a$.

This shows that 1d $\cN=4$ SQCD theories also
exhibit Seiberg-like duality which is very similar to
that of 2d, 3d and 4d \cite{Benini:2014mia,Benini:2011mf,Seiberg:1994pq}
with the same amount of supersymmetries. Of course, somewhat
 special feature of 1d version is that the theory experiences wall-crossing
so that the duality map changes when we go to $\zeta<0$ chamber of the original theory.
At this chamber, the JK-residue picks all poles from anti-fundamentals,
whose index is similarly mapped to the theory with $\zeta''>0$ chamber of
$U(N_a-N_c)$ gauge group. For this example, this selection merely
tells us whether $\mu^L$  or $\mu^R$ is the right mutation to perform.

The two types of dualities described above can be thought of as
prototypes of the right and left mutation of the quiver quantum mechanics
respectively.
One might expect that the mutation invariance for general quivers
can be straightforwardly proven by gauging the flavor nodes of 1d SQCD example,
but this procedure cannot be easily justified when degenerate singularities
appear. For such a case, the JK-residue description requires particular
order of taking residue integral, which makes it illegal to integrate out Cartans of
the mutating node prior to that of the flavorized nodes.
Despite these subtleties, there exist some classes of non-Abelian examples that
we can prove the mutation equivalence based on SQCD example.
Consider quivers with dimension vector $(1,1,N)$ and their sub-chambers
where $\zeta_3$ is right-most among  $\zeta$'s.
This corresponds to the three sub-chambers in Figure \ref{sub}, which belongs to
parts of I and IV physical chamber.

\begin{figure}[t]
\centering
\includegraphics[width=0.6\textwidth]{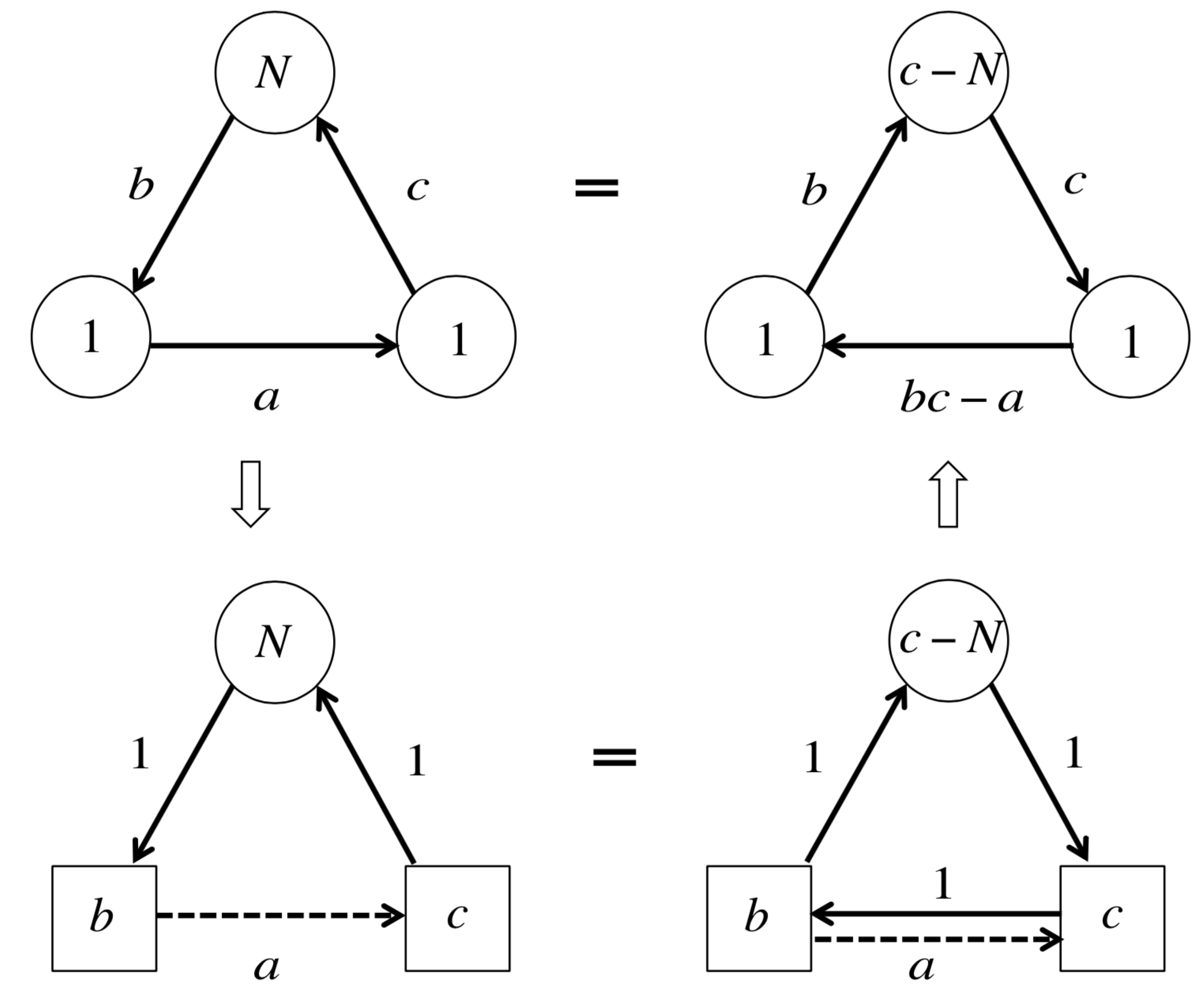}
\caption{\small  A diagrammatic proof of mutation invariance under
the right mutation $\mu^{R}$ on node 3 for $(1,1,N)$ quivers when
$\zeta_3 $ is right-most. The down arrow corresponds replacing two
Abelian nodes as $b$ and $c$ flavor nodes, with all chemical
potentials turned on. Dashed arrows, which are actually singlets
under $SU(b)\times SU(c)$, are the original bifundamentals
between node 1 and 2, and does not participate the mutation process.
They should be understood as singlet under $b$ and $c$ flavor nodes.
The up arrow is a reverse process of turning
off the chemical potentials and gauging the overall $U(1)$ in each
node. The equality between to the two bottom quivers follows from the
SQCD mutation.
 }
\label{proof}
\end{figure}

First of all, with respect to the Cartan directions,
$\{e_0;e_1,\cdots,e_N\}$, let us denote the
charges of three bifundamentals as
\begin{equation}
Q_X = e_0,~~Q_{Y_i}= e_i,~~Q_{Z_i}= -e_0-e_i\ ,
\end{equation}
and assign the $R$-charges by $(R_X,R_Y,R_Z)=(0,0,2)$ respectively,
which is consistent with a cubic superpotential of type $XYZ$.
Recall that, since only $R$-charge information enters the Witten index,
genericity of the superpotential consistent with $U(1)_R$ and
the gauge symmetry is implicitly assumed.
Then the one-loop determinant of $(1,1,N)$ quiver is given by \cite{HKY}
\begin{eqnarray}\nonumber
g&=&\left(\frac{1}{2\sinh[z/2]}\right)^{N+1}\prod_{i\neq j}
\frac{\sinh[(u_i-u_j)/2]}{\sinh[(u_i-u_j-z)/2]}\times \left(\frac{-\sinh[(u_0-z)/2]}{\sinh[u_0/2]}\right)^a\\
&&\times \prod_{i=1}^N\left(\frac{-\sinh[(u_i-z)/2]}{\sinh[u_i/2]}\right)^c
\prod_{i=1}^N\left(\frac{-\sinh[(-u_0-u_i)/2]}{\sinh[(-u_0-u_i+z)/2]}\right)^b\ .
\end{eqnarray}
If we put the $\eta$ parameter as
\begin{equation}
\eta = \zeta +\epsilon(Ne_1+(N-1)e_2+\cdots + 2e_{N-1}+e_N)\ ,
\end{equation}
with sufficiently small $\epsilon$ so that $\zeta$ and $\eta$
are in the same chamber in the space of charge vectors.

The JK-residue formula at each chamber reads as follow.
In chamber I, the index gets contribution
from poles of $X$ and $Y_i$'s, where we have
\begin{equation}
\Omega({\text{I}})=\frac{1}{N!}~\text{res}_{u_0=0}\text{res}_{u_N=0}\cdots \text{res}_{u_1=0}~g(u,z)\ .
\end{equation}
In chamber II, since $\eta$ is in a positive cone of $X$ and $Z_i$'s,
we have
\begin{equation}
\Omega({\text{II}})=\frac{(-1)^N}{N!}~\text{res}_{u_0=0}\text{res}_{u_N=z-u_0}\cdots\text{res}_{u_1=z-u_0} ~g(u,z).
\end{equation}
On the other hand, at chamber III and IV where $Y_i$'s and $Z_i$'s contribute,
the singularity is degenerate. A single ordered charge set
contributes to the integral at each chamber, which reads,
for chamber III,
\begin{equation}
\Omega({\text{III}})=\frac{(-1)^{N+1}}{N!}~
\text{res}_{u_0=z}\text{res}_{u_N=z-u_0}\cdots\text{res}_{u_1=z-u_0}~g(u,z)\ ,
\end{equation}
and for chamber IV, we have
\begin{equation}
\Omega({\text{IV}})=\frac{-1}{N!}~
\text{res}_{u_0=z}\text{res}_{u_N=0}\cdots\text{res}_{u_1=0}~g(u,z)\ .
\end{equation}
Note that the order of taking residue is crucial for the latter two cases.

Now, let us define new functions $G_1(u_0)$ and $G_2(u_0)$ as follows
\begin{equation}
G_1(u_0) := \frac{1}{N!}~\text{res}_{u_N=0}\cdots \text{res}_{u_1=0}~g(u,z)\ ,
\end{equation}
and
\begin{equation}
G_2(u_0) := \frac{(-1)^N}{N!}~\text{res}_{u_N=z-u_0}\cdots \text{res}_{u_1=z-u_0}~g(u,z)\ .
\end{equation}
Then For chamber I and IV, the index is expressed as
\begin{equation}
\Omega({\text{I}})=\text{res}_{u_0=0} G_1(u_0)\ ,
~~\text{and}~~\Omega({\text{IV}})=-\text{res}_{u_0=z} G_1(u_0)\ ,
\end{equation}
while for chamber II and III, we have
\begin{equation}
\Omega({\text{II}})= \text{res}_{u_0=0} G_2(u_0)\ ,
~~\text{and}~~\Omega({\text{III}})=-\text{res}_{u_0=z} G_2(u_0)\ .
\end{equation}
Meanwhile, for the dual quiver with ranks $(1,1,c-N)$ and intersection
numbers $(-b,-c,-bc+a)$, we similarly have
\begin{eqnarray}\nonumber
g_{\text{dual}}&=&\left(\frac{1}{2\sinh[z/2]}\right)^{c-N+1}\prod_{a\neq b}
\frac{\sinh[(v_a-v_b)/2]}{\sinh[(v_a-v_b-z)/2]}\times \left(\frac{-\sinh[(-v_0-z)/2]}{\sinh[-v_0/2]}\right)^{bc-a}\\
&&\times \prod_{a=1}^{c-N}\left(\frac{-\sinh[(-v_a-z)/2]}{\sinh[-v_a/2]}\right)^c
\prod_{a=1}^{c-N}\left(\frac{-\sinh[(v_0+v_a)/2]}{\sinh[(v_0+v_a+z)/2]}\right)^b\ ,
\end{eqnarray}
and the indices for the four chambers can be written as
\begin{eqnarray}\nonumber
&&\Omega({\widehat{\text{I}}})=-\text{res}_{v_0=0}\widehat{G}_1(v_0)\ ,~~~
\Omega({\widehat{\text{IV}}}) = \text{res}_{v_0=-z}\widehat{G}_1(v_0)\\
&&\Omega({\widehat{\text{II}}})=-\text{res}_{v_0=0}\widehat{G}_2(v_0)\ ,~~~
\Omega({\widehat{\text{III}}}) = \text{res}_{v_0=-z}\widehat{G}_2(v_0)\ ,
\end{eqnarray}
where
\begin{eqnarray}\nonumber
\widehat{G}_1(v_0)&=&\frac{(-1)^{c-N}}{(c-N)!}~\text{res}_{v_{c-N}=0}\cdots \text{res}_{v_{1}=0}
~g_{\text{dual}}(v_i,v_0)\ ,\\
\widehat{G}_2(v_0)&=&\frac{1}{(c-N)!}~\text{res}_{v_{c-N}=-v_0-z}\cdots \text{res}_{v_{1}=-v_0-z}
~g_{\text{dual}}(v_i,v_0)\ .
\end{eqnarray}
In order to prove the equivalence of the indices under the
right mutation of the node 3, we show below that
\begin{eqnarray}\label{show}
G_1(u_0)=\widehat{G}_2(u_0-z)
\end{eqnarray}
holds, from which it would follow that
\begin{equation}
\Omega({\text{I}}) =\Omega({\widehat{\rm III}}) \ ,~~~\Omega({\text{IV}}) =\Omega({\widehat{\rm II}}) \ ,
\end{equation}
where $R$-charges for the dual theory are now assigned as
$(R_{\widehat{X}},R_{\widehat{Y}},R_{\widehat{Z}})=(2,0,0)$.

For this purpose,
we introduce auxiliary variables $a_{\gamma=1,\cdots c}$ to
split the order $c$ pole defined by $u_i=0$ into sum over
residues over various simple poles;
\begin{eqnarray}\nonumber
G_1(u_0)&=&\frac{1}{N!}~\text{res}_{u_N=0}\cdots \text{res}_{u_1=0}~g(u,z)\\\nonumber
&=&\frac{1}{N!}~\text{res}_{u_N=0}\cdots \text{res}_{u_1=0}
\lim_{a_\gamma\rightarrow 0}~\tilde{g}(u,z,a_\gamma)\\\label{mmm}
&=&\frac{1}{N!} \lim_{a_\gamma\rightarrow 0} \sum_{\tau}
\text{res}_{u_N=a_{\tau(N)}}\cdots
\text{res}_{u_1=a_{\tau(1)}} ~\tilde{g}(u,z,a_\gamma)\ ,
\end{eqnarray}
where $\tilde{g}(u,z,a_\gamma)$ is defined by
\begin{eqnarray}\nonumber
\tilde{g}(u,z,a_\gamma)&=&\left(\frac{1}{2\sinh[z/2]}\right)^{N+1}\prod_{i\neq j}
\frac{\sinh[(u_i-u_j)/2]}{\sinh[(u_i-u_j-z)/2]}\times \left(\frac{-\sinh[(u_0-z)/2]}{\sinh[u_0/2]}\right)^a\\
&&\times \prod_{i=1}^N\prod_{\gamma=1}^c\frac{-\sinh[(u_i-a_\gamma-z)/2]}{\sinh[(u_i-a_\gamma)/2]}
\prod_{i=1}^N\left(\frac{-\sinh[(-u_0-u_i)/2]}{\sinh[(-u_0-u_i+z)/2]}\right)^b\ .
\end{eqnarray}
In the last line of (\ref{mmm}), the summation is taken over all different $c^N$ choices of
$\tau(i)$ for each Cartan $u_i$.
The evaluation of the residue integral for $G_1(u_0)$ becomes,
\begin{eqnarray}\nonumber
G_1(u_0)&=&\frac{1}{N!} \lim_{a_\gamma\rightarrow 0} \sum_{\tau}
\text{res}_{u_N=a_{\tau(N)}}\cdots
\text{res}_{u_1=a_{\tau(1)}} ~\tilde{g}(u,z,a_\gamma)\\\nonumber
&=&\lim_{a_\gamma\rightarrow 0}~ \frac{1}{2\sinh[z/2]}~
\sum_{C(c,N)}\prod_{\substack{\gamma\in A\\\gamma'\in A'}}
\frac{-\sinh[(a_\gamma-a_{\gamma'}-z)/2]}{\sinh[(a_\gamma-a_{\gamma'})/2]}
\left(\frac{-\sinh[(u_0-z)/2]}{\sinh[u_0/2]}\right)^a\\\label{arre}
&&\phantom{aaaaaaaaaaaaaaaaa}\times
\prod_{\gamma \in A}\left(\frac{-\sinh[(-u_0-a_\gamma)/2]}{\sinh[(-u_0-a_\gamma+z)/2]}\right)^b\ ,
\end{eqnarray}
where $C(c,N)$ denotes all combinations of subset $A=\{\gamma_i|~u_i=a_{\gamma_i},~i=1,\cdots N$
such that $a_{\gamma_i}\neq a_{\gamma_j}$ for all $i\neq j\}$, and $A'$
is complement of a set $A$.
Furthermore, using
\begin{eqnarray}
&&\left(\frac{-\sinh[(u_0-z)/2]}{\sinh[u_0/2]}\right)^a \cr\cr
&&=
\lim_{a_\gamma\rightarrow 0}
\left(\frac{-\sinh[-u_0/2]}{\sinh[(-u_0+z)/2]}\right)^{bc-a}
\left(\prod_{\gamma\in A\cup A'} \frac{-\sinh[(-u_0-a_\gamma+ z)/2]}{\sinh[(-u_0-a_\gamma)/2]}\right)^b\ ,
\end{eqnarray}
we can alternatively express each term in the sum of (\ref{arre}) as
\begin{eqnarray}
&&\prod_{\substack{\gamma\in A\\\gamma'\in A'}}
\frac{-\sinh[(a_\gamma-a_{\gamma'}-z)/2]}{\sinh[(a_\gamma-a_{\gamma'})/2]}\cr\cr
&& \times\left(\frac{-\sinh[-u_0/2]}{\sinh[(-u_0+z)/2]}\right)^{bc-a}
\left(\prod_{\gamma\in A'} \frac{-\sinh[(-u_0-a_\gamma+ z)/2]}{\sinh[(-u_0-a_\gamma)/2]}\right)^b\ ,
\end{eqnarray}
which is exactly equal to a term in $\widehat{G}_2(u_0-z)$.

Although the limit $a_\gamma\rightarrow 0$ is not well-defined for individual terms,
it can be shown that the limit gives finite answer when we sum up all combinations $C(c,N)$.
After we add up all terms in the summation, (\ref{arre}) can be written in a following form.
$$\frac{f(A_\gamma, U_0)}{\prod_{\gamma<\delta}(A_\gamma-A_\delta)
\prod_{\text{all }\gamma}(A_\gamma-U_0)}\ ,$$
where $U_0=e^{u_0/2}$, $A_\gamma=e^{a_\gamma/2}$, and $f(A_\gamma, U_0)$ is
an anti-symmetric polynomial in $A_\gamma$'s. Since every antisymmetric polynomial
is divisible by the Vandermonde determinant to a symmetric polynomial, the first factor
in the denominator is always canceled and $a_\gamma\rightarrow 0$ limit is well-defined
at the generic value of $U_0$. This gives
\begin{eqnarray}
\Omega({\text{I}}) =\Omega({\widehat{\text{III}})} ,&&\text{for region }(2_-,1_{\pm},3_+)\text{ and } (1_-,2_-,3_+)\cr
\Omega({\text{IV}}) =\Omega({\widehat{\text{II}})} ,&&\text{for region }(1_-,2_+,3_+)
\end{eqnarray}
under $\mu_3^R$ as promised, where each regions of original FI parameters are drawn in Figure \ref{sub}.

The left-mutation $\mu_3^L$  on node 3 when $\zeta_3$ is left-most can be
checked similarly, resulting in the identities
\begin{eqnarray}
\Omega({\text{II}}) =\Omega({\widehat{\text{IV}}}) ,&&\text{for region }(3_-,1_+,2_+)\text{ and } (3_-,2_\pm,1_+)\cr
\Omega({\text{III}}) =\Omega({\widehat{\text{I}}}) ,&&\text{for region }(3_-,1_-,2_+)
\end{eqnarray}
where we remind readers that the mutated quiver under $\mu_3^L$ is
not the same as the mutated quiver under $\mu_3^R$ unless $b=c$.

Let us briefly comment on what happens for the mutation procedure
when we start with a different $R$-charge assignment. Suppose we had
a triangle quiver with potential
$W=(XYZ)^2$, which requires $R_X+R_Y+R_Z=1$.
As can be inferred from (\ref{dualf}), the dual quiver has
new chiral fields $X',Y',Z'$ with $R$-charges $(R_X', R_Y',R_X+R_Y)$
where $R_X'+R_Y'=2-R_X-R_Y$,  as well as the original chiral $X$.
Because of $R$-charge mismatch, $X$ and $X'$ cannot form a mass
term $XX'$. The superpotential of the mutated theory is generic
of type $W=X'Y'Z'+(XX')^2$.\footnote{In standard mathematics literature,
such possibilities are precluded by assuming absence of 1-cycles
and 2-cycles \cite{Cluster,Keller}.}
\begin{figure}[h]
\centering
\includegraphics[width=0.7\textwidth]{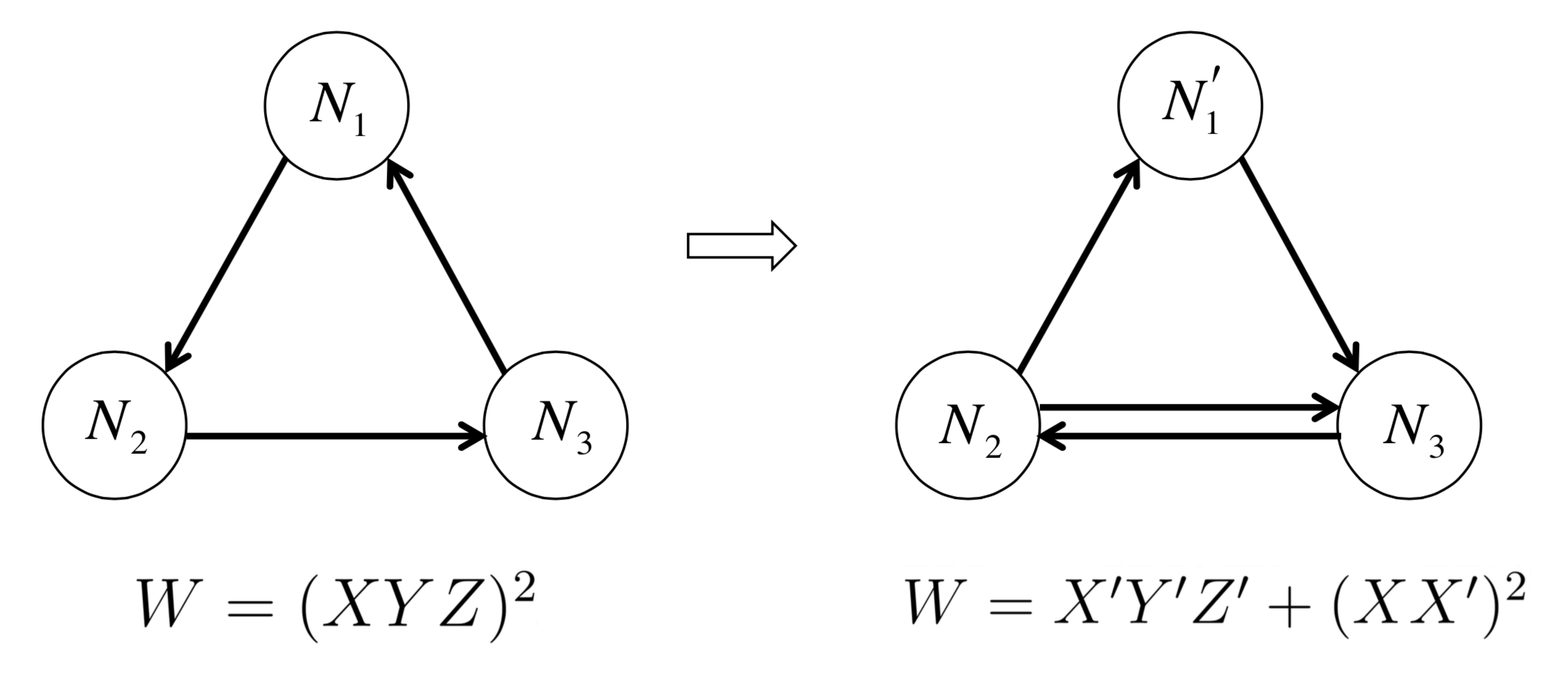}
\caption{Mutation of a quiver with non-generic superpotential}
\end{figure}

\section{Mutation $\tilde \mu_k$ on Quiver Invariant $\Omega_{\rm Q}\vert_{\rm Inv}$}\label{triangle}

We start with the observation that the alternate  mutation
rules $\tilde \mu^{L,R}$ preserve the quiver invariant regardless
of the choices of the chamber or the node, while, as we
saw in previous section, $\mu^{L,R}$ preserve Witten index of
individual chambers when the node is carefully chosen. Let us
start with the explanation of why $\mu_k$ cannot preserve
the quiver invariants.

Recall that the ordinary mutation rule $\mu$ preserves the
total charge $\Gamma=\mu_k(\Gamma)$ and, as we demonstrated
in previous section, also  preserves the index when $k$
and $L,R$ is appropriately chosen. For preservation of the
index
\begin{equation}
\Omega_{\rm Q}(\zeta) = \Omega_{\mu_k^{L,R}({\rm Q})}(\mu_k^{L,R}(\zeta))\ ,
\end{equation}
the allowed choice of the mutation node $k$ and the
choice between $L$ and $R$ are severely restricted by $\zeta$.
The choice becomes clearer when in a given chamber  we can
take a pair, $\zeta_k$ and $\zeta_{k'}$, large positive and
large negative, respectively and relative to other $\zeta_l$'s:
One must perform either $\mu_k^L$ or $\mu_{k'}^R$.

Therefore, a chamber of ${\rm Q}$ is mapped to a chamber of such
a mutated quiver $\mu_k({\rm Q})$ but another chamber of ${\rm Q}$
is not necessarily mapped to another chamber of the same $\mu_k({\rm Q})$.
Instead, the latter would be generically mapped to a chamber of
a differently mutated quiver $\mu_l({\rm Q})$. Generally, the
number of chambers for ${\rm Q}$ is not necessarily the same as that
of $\mu_k({\rm Q})$ for a given $k$, so wall-crossing pattern
of ${\rm Q}$ cannot be the same as that of $\mu_k({\rm Q})$ anyhow.
It follows that quiver invariant of $Q$ cannot be generally
the same as that of $\mu_k({\rm Q})$, even though the two
share a chamber with the same moduli space topology somewhere
in the respective FI constant space.

This happens because the quiver invariant  is not a property
of the quiver moduli spaces but rather of the quiver itself,
or of the gauged quantum mechanics as a whole.
On the other hand, the mutation map originates in mathematics
literature as a transformation of the quiver diagrams themselves,
and it would be strange if there is no definite behavior of
the quiver invariants under the mutation maps.

A very strong hint of how quiver invariant should behave
under mutation \cite{Manschot:2013dua}
is found in the partition sum expansion of
Eq.~(\ref{MPS}). The formula implies that quiver
invariants of all subquivers should behave like an ``elementary"
node in the induced quiver ${\rm Q}/\{{\rm Q}_p\}$ for any given partition
${\rm Q}=\oplus {\rm Q}_p$. So as far as  ${\rm Q}/\{{\rm Q}_p\}$ goes, the
subquiver ${\rm Q}_p$ of total charge $\Gamma_p=\sum_i N_i^{(p)}\gamma_i$
must behave as if it is an elementary node. Therefore, if
this expansion makes sense, $\Gamma_p$ must be mapped exactly
as if $\Gamma_p$ belongs to a single node of quiver.
This is precisely the mutation maps $\tilde\mu$ introduced
earlier. Since any quiver can be a  subquiver of infinite number
of larger quivers, this means that the mutation preserving
the quiver invariant has to be the modified one $\tilde\mu$
\begin{equation}
\Omega_{\rm Q}\biggl\vert_{\rm Inv} = \Omega_{\tilde \mu_k({\rm Q})}\biggl\vert_{\rm Inv}\ ,
\end{equation}
which brings us to the assertion at the head of this section.

Recall that the action of $\tilde \mu$ is the same as $\mu$
except for the action on the ranks $N_i$, as in Eq.~(\ref{tmuN}).
A bonus is that since the quiver invariant is a chamber-independent
concept, we need not be careful about $\zeta$'s, and the
restriction on the choice of $k$ does not exist. Therefore,
the above holds for any choice of $k$,
\begin{equation}
\Omega_{\rm Q}\biggl\vert_{\rm Inv} = \Omega_{\tilde \mu_k({\rm Q})}\biggl\vert_{\rm Inv},\qquad \hbox{for all } k
\end{equation}
and, for a given quiver with $K$ nodes, one finds as many as
$K$ mutated quivers, upon a single mutation step, that shares
the same quiver invariant. This is in fact much more powerful
and useful statement than the invariance of Witten index under
ordinary mutation $\mu$. The latter is mired by the rather
complicated choice of the node to be mutated, while the
invariance of the quiver invariants under $\tilde \mu$ is
completely independent of FI constants.

For example, let $\Omega^{N,m,n}_{a,b,c}(\zeta) $'s be
the indices for a cyclic triangle quiver of ranks $N,m,n$
with the opposing intersection numbers $a,b, c$. Using
$\tilde \mu$ mutation on the first node, we obtain
$$\Omega^{1,1,1}_{a,b,c}\biggl\vert_{\rm Inv}
= \Omega^{{\rm min}[b,c]-1,1,1}_{a-bc,-b,-c}\biggl\vert_{\rm Inv} \ . $$
For a non-Abelian cyclic triangle, the same procedure gives us
$$\Omega^{N,m,n}_{a,b,c}\biggl\vert_{\rm Inv}
= \Omega^{{\rm min}[bn,cm]-N,m,n}_{a-bc,-b,-c}\biggl\vert_{\rm Inv}\ .$$
Negative intersection number means flipping of arrows relative
to the original quiver, but the overall direction does not matter
so we will sometimes flip the intersection numbers altogether.

Appendix will discuss the validity of this claim for a series of
$(N,1,1)$ quivers which are obtained via $\tilde\mu$ mutation from
Abelian triangle quivers. Here we simply list the explicit forms
of the relevant quiver invariants:

\begin{eqnarray}\label{2kk}
&&\Omega^{k-1,1,1}_{k^2-2,k,k}\biggl\vert_{\rm Inv}=\Omega^{1,1,1}_{2,k,k}\biggl\vert_{\rm Inv}=k-1 \ ,
\end{eqnarray}

\begin{eqnarray}\label{3kk}
&&\Omega^{k-1,1,1}_{k^2-3,k,k}\biggl\vert_{\rm Inv}
=\Omega^{1,1,2}_{3,k,2k}\biggl\vert_{\rm Inv}=\Omega^{1,1,1}_{3,k,k}\biggl\vert_{\rm Inv}=\left[\frac{(k-1)(k-2)}{2} \right]\times ({\bf y}+1/{\bf y}) \ ,
\end{eqnarray}

\begin{eqnarray}\label{3kk+1}
&&\Omega^{k-1,1,1}_{k(k+1)-3,k,k+1}\biggl\vert_{\rm Inv}=\Omega^{1,2,1}_{3,2k+3,k+1}\biggl\vert_{\rm Inv}
=\Omega^{1,1,2}_{3,k,2k-1}\biggl\vert_{\rm Inv}=\Omega^{1,1,1}_{3,k,k+1}\biggl\vert_{\rm Inv}
=\frac{(k-1)(k+2)}{2}\ ,  \cr &&
\end{eqnarray}

\begin{eqnarray}\label{4kk}
&&\Omega^{k,1,1}_{k^2-4,k,k}\biggl\vert_{\rm Inv} =
\Omega^{1,3,1}_{4,3k,k}\biggl\vert_{\rm Inv} =
\Omega^{1,1,1}_{4,k,k}\biggl\vert_{\rm Inv} \cr\cr\cr
&=&\frac{(k-1)(k-2)(k-3)}{6}\left({\bf y}^2+1+1/{\bf y}^2\right)+ \frac{(k-2)(k^2+1)}{2} \ ,
\end{eqnarray}

\begin{eqnarray}\label{4kk+1}
&&\Omega^{k-1,1,1}_{k^2+k-4,k,k+1}\biggl\vert_{\rm Inv} =
\Omega^{1,3,1}_{4,3k+2,k+1}\biggl\vert_{\rm Inv} =
\Omega^{1,1,3}_{4,k,3k-1}\biggl\vert_{\rm Inv} =
\Omega^{1,1,1}_{4,k,k+1}\biggl\vert_{\rm Inv} \cr\cr\cr
&=&\frac{(k - 1)(k - 2)(2 k + 3)}{6}\left({\bf y}+1/{\bf y}\right)
\end{eqnarray}

\begin{eqnarray}\label{4kk+2}
&&\Omega^{k-1,1,1}_{k^2+2k-4,k,k+2}\biggl\vert_{\rm Inv} =
\Omega^{1,3,1}_{4,3k+8,k+2}\biggl\vert_{\rm Inv} =
\Omega^{1,1,3}_{4,k,3k-2}\biggl\vert_{\rm Inv} =
\Omega^{1,1,1}_{4,k,k+2}\biggl\vert_{\rm Inv} \cr\cr\cr
&=& \frac{(k - 1)(k^2 + 4k + 6)}{6}
\end{eqnarray}

This generalizes to general quivers as follows. Given a quiver,
let us concentrate on a mutating node, say of rank $N$, and nodes of
rank $m_i$ and $n_p$, connected to it by, respectively, $c_i$
ingoing or $b_p$ outgoing arrows. The index may be denoted as
$$\Omega^{N,m_i,n_p,\dots}_{a_{ij};a_{ip};a_{pq},b_p,c_i,\dots}(\vec \zeta)\ ,$$
where $b$'s and $c$'s denote, respectively, ingoing and outgoing
intersection numbers, all taken to be positive, from the mutating
node. The three set of numbers encoded in the three matrices, $a_{ij};a_{ip};a_{pq}$,
are intersection numbers among the nodes connected to the mutation node
in the initial quiver.
With this, the mutation rule for the quiver invariant is
$$\Omega^{N,m_i,n_p,\dots}_{a_{ij};a_{ip};a_{pq},b_p,c_i,\dots}\biggl\vert_{\rm Inv}=
\Omega^{min[b\cdot n,c\cdot m]-N,m_i,n_p,\dots}_{a_{ij};a_{ip}-b_pc_i;a_{pq},-b_p,-c_i,\dots}\biggl\vert_{\rm Inv} \ . $$

As we emphasized already, the quiver invariants are
properties of the quivers themselves and therefore we
do not need to be selective in choosing mutation nodes.
Generally, given a quiver with $K$ number of nodes,
there are as many as $K$ mutated quivers whose quiver
invariants all agree with the quiver invariant of the
original quiver. With such a strong and universal statement,
a very tantalizing question that should be explored further
is whether this notion of quiver invariants and their
invariance under $\tilde\mu$ mutations is hidden in
the existing cluster algebra structure of quivers, or
can be embedded into its generalization.

\section{Summary}

In this note, we have explored how mutation maps of
quiver diagram work to preserve Witten indices and
quiver invariants, relying on prototypical examples
of triangle quivers.

For Witten indices of 1d quiver theories, which are
chamber-dependent quantities, mutations are far more
restricted than its higher dimensional counterpart.
For any given point in $\zeta$ space, only two possible
mutations exist, which divides physical chambers further
into sub-chambers. The allowed mutation maps a sub-chamber
into a physical chamber of the mutated quiver, while
disallowed mutation actually fails preserve the Witten index.
This identifies a specific chamber of quiver with a specific
chamber of the mutated quiver. We have shown how this
equality is realized at the level of Witten index
expressed as residue integrals for simple class of $(1,1,N)$
triangle quivers.

Quiver invariants, on the other hand, is an intrinsic
quantity of the quiver itself rather than its chambers.
As such, the complicated (sub-)chamber-dependence should
be unnecessary, and we argued that any given node can
be mutated to give another quiver of the same quiver
invariant. The mutation rule $\tilde \mu$ for them
differs slightly from those $\mu$'s used for the Witten
indices, in that $\tilde\mu$ acts differently on the rank
vectors than $\mu$. With a single step of mutation,
a quiver with $K$ nodes is mapped to $K$ $\tilde\mu$-mutated
quivers, therefore. Again we have tested this assertion
for the simple classes of triangle quivers, by explicit
computations.

\vskip 1cm
\centerline{\bf Acknowledgement}
\vskip 5mm\noindent
We are indebted to Kentaro Hori and Zhao-Long Wang for
useful conversations, and Kyungyong Lee for extensive
discussions of the Cluster Algebra. H.K. and S-J.L. are
grateful to Korea Institute for Advanced Study for
hospitality.
The work of S.-J.L. is supported in part by NSF grant PHY-1417316. H.K. was supported
by the Perimeter Institute for Theoretical Physics. Research at Perimeter Institute
is supported by the Government of Canada through Industry Canada and by the
Province of Ontario through the Ministry of Economic Development and Innovation.
The work of H.K. was made possible through the support of a grant from John Templeton
Foundation. The opinions expressed in this publication are those of the author and
do not necessarily reflect the views of the John Templeton Foundation.

\appendix
\section{Quiver Invariants from Direct Index Computations}\label{appA}

We list indices in the four chambers of $(N,1,1)$ quivers, obtained
by $\tilde \mu$ mutation from $(1,1,1)$ quivers, computed
with the help of HKY routine. This class of quivers comes up to
four different chambers in the FI space, and we display index
for each chamber. The last item for each quiver is the quiver
invariant, extracted by comparing these indices against the
Coulombic computations of MPS.\footnote{The Coulombic computation,
which we do not explicitly display here, were obtained using
the mathematica codes supplied by Manschot et.al. together
with Ref.~\cite{Manschot:2013sya}.  We gratefully acknowledge
their generosity for making the code public. }

In the MPS expansion,
quiver invariants of all subquivers are left as unknown input
parameters, so that the comparison against HKY computation
fixes these quantity. For the MPS expansion, for which a mathematica
package supplied in Ref.~\cite{Manschot:2013sya} is used, the quantity being computed is
usually Poincare polynomial rather than the index. However,
the expansion formula itself should be applicable to the index
as well, because purely Coulombic states are neutral under $U(1)_R$
while the states counted by quiver invariants are neutral
under $SU(2)_R$. The actual vacua are obtained by simple
tensor product of these two classes of states, and their
quantum numbers are already manifest individually.

The following confirms the quiver invariants of all $(2,1,1)$
quivers that appear in Eq.~(\ref{3kk}), up to $k=9$, and
Eq.~(\ref{3kk+1}), up to $k=8$,
by direct computations. To extract quiver invariant of given
$(N,1,1)$ quiver, by comparing HKY index against MPS's partition
sum, one ends up computing quiver invariants of $(n,1,1)$
quivers for all $n\le N$ recursively. However, for all of
examples below, $(1,1,1)$ quivers happen to carry no nontrivial
quiver invariant. For Abelian cyclic quivers, the geometric
characterization of the quiver invariant in
Refs.~\cite{Lee:2012sc, Lee:2012naa} is applicable,
so that the quiver invariant is null whenever
there is a chamber of null Higgs moduli space. For this reason,
we chose not to display the indices of the Abelian version.

\begin{itemize}

\item
$(2,1,1)$-Quiver with intersection numbers $(2k,k,3)$

\begin{eqnarray}
&&\Omega^{2,1,1}_{6,3,3}=\left(\begin{array}{l}
0 \\
0 \\
0 \\
0
\end{array}\right. \cr\cr
&&\Omega^{2,1,1}_{6,3,3}\biggl\vert_{\rm Inv} =1/{\bf y}+{\bf y}
\cr\cr\cr
&&\Omega^{2,1,1}_{8,4,3}=\left(\begin{array}{l}
-1/{\bf y}^5-2/{\bf y}^3-1/{\bf y}-{\bf y}-2{\bf y}^3-{\bf y}^5 \\
2/{\bf y}+2{\bf y} \\
2/{\bf y}+2{\bf y}\\
2/{\bf y}+2{\bf y}
\end{array}\right. \cr\cr
&&\Omega^{2,1,1}_{8,4,3}\biggl\vert_{\rm Inv} =3/{\bf y}+3{\bf y}
\cr\cr\cr
&&\Omega^{2,1,1}_{10,5,3}=\left(\begin{array}{l}
 -1/{\bf y}^9-2/{\bf y}^7-4/{\bf y}^5-6/{\bf y}^3-2/{\bf y}-2{\bf y}-6{\bf y}^3-4{\bf y}^5-2{\bf y}^7-{\bf y}^9 \\
5/{\bf y}+5{\bf y}\\
5/{\bf y}+5{\bf y}\\
5/{\bf y}+5{\bf y}
\end{array}\right. \cr\cr
&&\Omega^{2,1,1}_{10,5,3}\biggl\vert_{\rm Inv} =6/{\bf y}+6{\bf y}
\cr\cr\cr
&&\Omega^{2,1,1}_{12,6,3}=\left(\begin{array}{l}
 -1/{\bf y}^{13}-2/{\bf y}^{11}-4/{\bf y}^9-6/{\bf y}^7-9/{\bf y}^5-11/{\bf y}^3
 \\ \hskip 2cm -3/{\bf y}-3{\bf y}-11{\bf y}^3-9{\bf y}^5-6{\bf y}^7-4{\bf y}^9-2{\bf y}^{11}-{\bf y}^{13} \\
9/{\bf y}+9{\bf y} \\
9/{\bf y}+9{\bf y} \\
9/{\bf y}+9{\bf y}
\end{array}\right. \cr\cr
&&\Omega^{2,1,1}_{12,6,3}\biggl\vert_{\rm Inv} =10/{\bf y}+10{\bf y}
\cr\cr\cr
&&\Omega^{2,1,1}_{14,7,3}=\left(\begin{array}{l}
-1/{\bf y}^{17}-2/{\bf y}^{15}-4/{\bf y}^{13}-6/{\bf y}^{11}-9/{\bf y}^9-12/{\bf y}^7
 \\ \hskip 2cm -15/{\bf y}^5-17/{\bf y}^3-4/{\bf y}-4{\bf y}-17{\bf y}^3-15{\bf y}^5
  \\ \hskip 4cm -12{\bf y}^7-9{\bf y}^9-6{\bf y}^{11}-4{\bf y}^{13}-2{\bf y}^{15}-{\bf y}^{17} \\
14/{\bf y}+14{\bf y} \\
14/{\bf y}+14{\bf y} \\
14/{\bf y}+14{\bf y}
\end{array}\right. \cr\cr
&&\Omega^{2,1,1}_{14,7,3}\biggl\vert_{\rm Inv} =15/{\bf y}+15{\bf y}
\cr\cr\cr
&&\Omega^{2,1,1}_{16,8,3}=\left(\begin{array}{l}
 -1/{\bf y}^{21}-2/{\bf y}^{19}-4/{\bf y}^{17}-6/{\bf y}^{15}-9/{\bf y}^{13}-12/{\bf y}^{11}-16/{\bf y}^9
  \\ \hskip 2cm-19/{\bf y}^7-22/{\bf y}^5-24/{\bf y}^3-5/{\bf y}-5 {\bf y}-24 {\bf y}^3 -22 {\bf y}^5-19 {\bf y}^7
  \\ \hskip 4cm -16 {\bf y}^9-12 {\bf y}^{11}-9 {\bf y}^{13}-6 {\bf y}^{15}-4 {\bf y}^{17}-2 {\bf y}^{19}-{\bf y}^{21} \\
20/{\bf y}+20{\bf y} \\
20/{\bf y}+20{\bf y} \\
20/{\bf y}+20{\bf y}
\end{array}\right. \cr\cr
&&\Omega^{2,1,1}_{16,8,3}\biggl\vert_{\rm Inv} =21/{\bf y}+21{\bf y}
\cr\cr\cr
&&\Omega^{2,1,1}_{18,9,3}=\left(\begin{array}{l}
-1/{\bf y}^{25}-2/{\bf y}^{23}-4/{\bf y}^{21}-6/{\bf y}^{19}-9/{\bf y}^{17}-12/{\bf y}^{15}
  \\ \hskip 2cm -16/{\bf y}^{13}-20/{\bf y}^{11}-24/{\bf y}^9-27/{\bf y}^7-30/{\bf y}^5-32/{\bf y}^3-6/{\bf y}
  \\ \hskip 4cm -6 {\bf y}-32 {\bf y}^3 -30 {\bf y}^5-27 {\bf y}^7-24 {\bf y}^9-20 {\bf y}^{11}-16 {\bf y}^{13}
  \\ \hskip 6cm -12 {\bf y}^{15}-9 {\bf y}^{17}-6 {\bf y}^{19}-4 {\bf y}^{21}-2 {\bf y}^{23}-{\bf y}^{25} \\
27/{\bf y}+27{\bf y} \\
27/{\bf y}+27{\bf y} \\
27/{\bf y}+27{\bf y}
\end{array}\right. \cr\cr
&&\Omega^{2,1,1}_{18,9,3}\biggl\vert_{\rm Inv} =28/{\bf y}+28{\bf y} \cr\cr\nonumber
\end{eqnarray}

\item
$(2,1,1)$-Quiver with intersection numbers $(2k+3,k+1,3)$

\begin{eqnarray}
&&\Omega^{2,1,1}_{9,4,3}=\left(\begin{array}{l}
1/{\bf y}^6+2/{\bf y}^4+4/{\bf y}^2+10+4{\bf y}^2+2{\bf y}^4+{\bf y}^6 \\
1/{\bf y}^2+7+{\bf y}^2 \\
6 \\
6
\end{array}\right. \cr\cr
&&\Omega^{2,1,1}_{9,4,3}\biggl\vert_{\rm Inv} =5
\cr\cr\cr\cr
&&\Omega^{2,1,1}_{11,5,3}=\left(\begin{array}{l}
1/{\bf y}^{10}+2/{\bf y}^8+4/{\bf y}^6+6/{\bf y}^4+8/{\bf y}^2+ 18+8{\bf y}^2+6{\bf y}^4+4{\bf y}^6+2{\bf y}^8+{\bf y}^{10} \\
1/{\bf y}^2 +11+{\bf y}^2 \\
10 \\
10
\end{array}\right.\cr\cr
&&\Omega^{2,1,1}_{11,5,3}\biggl\vert_{\rm Inv} =9
\cr\cr\cr\cr
&&\Omega^{2,1,1}_{13,6,3}=\left(\begin{array}{l}
1/{\bf y}^{14}+2/{\bf y}^{12}+4/{\bf y}^{10}+6/{\bf y}^8+9/{\bf y}^6+11/{\bf y}^4+13/{\bf y}^2
\\ \hskip 2cm +28 +13{\bf y}^2+11{\bf y}^4+9{\bf y}^6+6{\bf y}^8+4{\bf y}^{10}+2{\bf y}^{12}+{\bf y}^{14} \\
1/{\bf y}^2+16+{\bf y}^2 \\
15 \\
15
\end{array}\right.\cr\cr
&&\Omega^{2,1,1}_{13,6,3}\biggl\vert_{\rm Inv} =14
\cr\cr\cr\cr
&&\Omega^{2,1,1}_{15,7,3}=\left(\begin{array}{l}
1/{\bf y}^{18}+2/{\bf y}^{16}+4/{\bf y}^{14}+6/{\bf y}^{12}+9/{\bf y}^{10}+12/{\bf y}^8
\\ \hskip 2cm +15/{\bf y}^6+17/{\bf y}^4+19/{\bf y}^2+40+19{\bf y}^2+17{\bf y}^4+15{\bf y}^6
\\ \hskip 4cm +12{\bf y}^8+9{\bf y}^{10}+6{\bf y}^{12}+4{\bf y}^{14}+2{\bf y}^{16}+{\bf y}^{18} \\
1/{\bf y}^2+22+{\bf y}^2\\
21\\
21
\end{array}\right.\cr\cr
&&\Omega^{2,1,1}_{15,7,3}\biggl\vert_{\rm Inv} =20
\cr\cr\cr\cr
&&\Omega^{2,1,1}_{17,8,3}=\left(\begin{array}{l}
1/{\bf y}^{22}+2/{\bf y}^{20}+4/{\bf y}^{18}+6/{\bf y}^{16}+9/{\bf y}^{14}+12/{\bf y}^{12}+16/{\bf y}^{10} +19/{\bf y}^8
\\ \hskip 2cm+22/{\bf y}^6+24/{\bf y}^4+26/{\bf y}^2+54+26{\bf y}^2+24{\bf y}^4+22{\bf y}^6
\\ \hskip 4cm +19{\bf y}^8+16{\bf y}^{10}+12{\bf y}^{12}+9{\bf y}^{14}+6{\bf y}^{16}+4{\bf y}^{18}+2{\bf y}^{20}+{\bf y}^{22} \\
1/{\bf y}^2+29+{\bf y}^2\\
28\\
28
\end{array}\right.\cr\cr
&&\Omega^{2,1,1}_{17,8,3}\biggl\vert_{\rm Inv} =27
\cr\cr\cr\cr
&&\Omega^{2,1,1}_{19,9,3}=\left(\begin{array}{l}
1/{\bf y}^{26}+2/{\bf y}^{24}+4/{\bf y}^{22}+6/{\bf y}^{20}+9/{\bf y}^{18}+12/{\bf y}^{16}+16/{\bf y}^{14}
\\ \hskip 1.5cm +20/{\bf y}^{12}  +24/{\bf y}^{10}+27/{\bf y}^8 +30/{\bf y}^6+32/{\bf y}^4+34/{\bf y}^2
\\ \hskip 3cm +70+34 {\bf y}^2+32 {\bf y}^4+30 {\bf y}^6+27 {\bf y}^8+24 {\bf y}^{10} +20 {\bf y}^{12}
\\ \hskip 4.5cm +16 {\bf y}^{14}+12 {\bf y}^{16}+9 {\bf y}^{18}+6 {\bf y}^{20}+4 {\bf y}^{22}+2 {\bf y}^{24}+{\bf y}^{26}\\
1/{\bf y}^2 +37 +{\bf y}^2\\
36\\
36
\end{array}\right.\cr\cr
&&\Omega^{2,1,1}_{19,9,3}\biggl\vert_{\rm Inv} = 35 \cr\cr\nonumber
\end{eqnarray}

\item
$(2,1,1)$-Quiver with intersection numbers $(2k-1,k,3)$
\begin{eqnarray}
&&\Omega^{2,1,1}_{5,3,3}=\left(\begin{array}{l}
 6 \\
 6 \\
1/{\bf y}^2+7+{\bf y}^2 \\
1/{\bf y}^2 +7+{\bf y}^2
\end{array}\right. \cr\cr
&&\Omega^{2,1,1}_{5,3,3}\biggl\vert_{\rm Inv} =5
\cr\cr\cr\cr
&&\Omega^{2,1,1}_{7,4,3}=\left(\begin{array}{l}
1/{\bf y}^4+2/{\bf y}^2 +13+2{\bf y}^2+{\bf y}^4 \\
10 \\
1/{\bf y}^2 +11+{\bf y}^2 \\
1/{\bf y}^2 +11+{\bf y}^2
\end{array}\right.\cr\cr
&&\Omega^{2,1,1}_{7,4,3}\biggl\vert_{\rm Inv} =9
\cr\cr\cr\cr
&&\Omega^{2,1,1}_{9,5,3}=\left(\begin{array}{l}
1/{\bf y}^8+2/{\bf y}^6+4/{\bf y}^4+6/{\bf y}^2 +22+6{\bf y}^2+4{\bf y}^4+2{\bf y}^6+{\bf y}^8 \\
15 \\
1/{\bf y}^2 +16+{\bf y}^2 \\
1/{\bf y}^2 +16+{\bf y}^2
\end{array}\right.\cr\cr
&&\Omega^{2,1,1}_{9,5,3}\biggl\vert_{\rm Inv} =14
\cr\cr\cr\cr
&&\Omega^{2,1,1}_{11,6,3}=\left(\begin{array}{l}
1/{\bf y}^{12}+2/{\bf y}^{10}+4/{\bf y}^8+6/{\bf y}^6+9/{\bf y}^4+11/{\bf y}^2
\\ \hskip 2cm +33+11{\bf y}^2+9{\bf y}^4+6{\bf y}^6+4{\bf y}^8+2{\bf y}^{10}+{\bf y}^{12} \\
21 \\
1/{\bf y}^2 +22+{\bf y}^2\\
1/{\bf y}^2 +22+{\bf y}^2
\end{array}\right. \cr\cr
&&\Omega^{2,1,1}_{11,6,3}\biggl\vert_{\rm Inv} =20
\cr\cr\cr\cr
&&\Omega^{2,1,1}_{13,7,3}=\left(\begin{array}{l}
1/{\bf y}^{16}+2/{\bf y}^{14}+4/{\bf y}^{12}+6/{\bf y}^{10}+9/{\bf y}^8+12/{\bf y}^6+15/{\bf y}^4+17/{\bf y}^2
\\ \hskip 2cm +46+17{\bf y}^2+15{\bf y}^4+12{\bf y}^6+9{\bf y}^8+6{\bf y}^{10}+4{\bf y}^{12}+2{\bf y}^{14}+{\bf y}^{16} \\
28 \\
1/{\bf y}^2 +29+{\bf y}^2\\
1/{\bf y}^2 +29+{\bf y}^2
\end{array}\right. \cr\cr
&&\Omega^{2,1,1}_{13,7,3}\biggl\vert_{\rm Inv} =27
\cr\cr\cr\cr
&&\Omega^{2,1,1}_{15,8,3}=\left(\begin{array}{l}
1/{\bf y}^{20}+2/{\bf y}^{18}+4/{\bf y}^{16}+6/{\bf y}^{14}+9/{\bf y}^{12}+12/{\bf y}^{10}+16/{\bf y}^8
\\ \hskip 2cm +19/{\bf y}^6+22/{\bf y}^4+24/{\bf y}^2+61 +24{\bf y}^2+22{\bf y}^4+19{\bf y}^6
\\ \hskip 4cm +16{\bf y}^8+12{\bf y}^{10}+9{\bf y}^{12}+6{\bf y}^{14}+4{\bf y}^{16}+2{\bf y}^{18}+{\bf y}^{20} \\
36 \\
1/{\bf y}^2 +37+{\bf y}^2\\
1/{\bf y}^2 +37+{\bf y}^2
\end{array}\right. \cr\cr
&&\Omega^{2,1,1}_{15,8,3}\biggl\vert_{\rm Inv} =35\cr\cr\nonumber
\end{eqnarray}

\end{itemize}
In all examples above, the computed quiver invariants
agree with predictions from the mutations $\tilde \mu$,
displayed in Section~\ref{triangle}.

We have also confirmed the quiver invariants in
all $(3,1,1)$ quivers that appear in Eqs.~(\ref{4kk}-\ref{4kk+2})
up to $k=9$, via direct computations along the same
line as the  $(2,1,1)$ cases above.

\end{document}